\documentclass[traditabstract]{aa}

\usepackage{graphicx}
\usepackage{txfonts}
\usepackage{xcolor}
\usepackage{comment}
\usepackage{array}
\usepackage{natbib}
\bibpunct{(}{)}{;}{a}{}{,}
\usepackage{hyperref}
\hypersetup{colorlinks=true,
            citecolor=blue,
            linkcolor=blue}
\usepackage{color}
\newcommand{\orcid}[1]{\protect\href{https://orcid.org/#1}{\protect\includegraphics[width=8pt]{orcid.png}}}
\newcolumntype{H}{>{\setbox0=\hbox\bgroup}c<{\egroup}@{}}

\def\be {\begin{equation}}
\def\ee {\end{equation}}
\def\osdot {\star}
\def\mp {m}
\def\Rp {R}

\def\ms {M_\osdot}
\def\Eo {E}
\def\nG {G}
\def\vG {\vec{\nG}}
\def\nL {L}
\def\vL {\vec{\nL}}
\def\vT {\vec{T}}
\def\Gc {{\cal G}}
\def\vk {\vec{k}}
\def\vp {\vec{p}}
\def\vq {\vec{q}}
\def\vs {\vec{s}}
\def\vg {\vec{g}}
\def\bg {b_g}
\def\ba {b_a}
\def\VA {V_a}
\def\om {\omega}
\def\At {{\cal K}}

\def\vw {\vec{\om}}

\def\vr {\vec{r}}
\def\ur {\vec{\hat r}}

\def\ii {\mathrm{i}} 
\def\lv {\upsilon} 
\def \cT {x}
\def \HZT {X_k^{-3,0}}
\def \HDPT {X_k^{-3,2}}
\def \HDMT {X_k^{-3,-2}}
\def \HZD {X_k^{-2,0}}
\def \HDPD {X_k^{-2,2}}
\def \HDMD {X_k^{-2,-2}}
\def \ke {k_{\rm e}}
\def \kf {k_{\rm f}}
\def \taue {\tau_e}
\def \ta {\tau_A}
\def \tauv {\tau_v}

\def \sq {\sqrt{1-e^2}}

\def\figpath{}
\def \llabel#1{\label{#1}}
\begin{document}

\title{Tidal evolution of Earth-like planets in the habitable zone of low-mass stars}
\titlerunning{Tidal evolution of Earth-like planets in the habitable zone of low-mass stars}

\authorrunning{E. F. S. Valente et al.}

\author{
E. F. S. Valente\inst{1}
 \and
A. C. M. Correia\inst{1,2}
\and
P. Auclair-Desrotour\inst{2}
\and
M. Farhat\inst{2}
\and
J. Laskar\inst{2}
}

\institute{
CFisUC, Departamento de F\'isica, Universidade de Coimbra, 3004-516 Coimbra, Portugal
\and 
IMCCE, UMR8028 CNRS, Observatoire de Paris, PSL Universit\'e, 77 Av. Denfert-Rochereau, 75014 Paris, France
}

\date{\today; Received; accepted To be inserted later}

% \abstract{}{}{}{}{} 
% 5 {} token are mandatory
 
  \abstract{Earth-like planets in the habitable zone of low-mass stars undergo strong tidal effects that modify their spin states. These planets are expected to host dense atmospheres that can also play an important role in the spin evolution. On one hand, gravitational tides tend to synchronise the rotation with the orbital mean motion, but on the other hand, thermal atmospheric tides push the rotation away and may lead to asynchronous equilibria. Here, we investigate the complete tidal evolution of Earth-like planets by taking into account the effect of obliquity and eccentric orbits. We adopted an Andrade rheology for the gravitational tides and benchmarked the unknown parameters with the present rotation of Venus. %We find that the time exponent $\alpha = 0.3$ and a Maxwell relaxation time of $\tau \approx 1500$~yr better fit the observations. 
  We then applied our model to Earth-like planets, and we show that asynchronous rotation can be expected for planets orbiting stars with masses between 0.4 and 0.9~$M_\odot$ and semi-major axes between 0.2 and 0.7~au. Interestingly, we find that Earth-like planets in the habitable zone of stars with masses $\sim 0.8$~$M_\odot$ may end up with an equilibrium rotation of 24~hours. We additionally find that these planets can also develop high obliquities, which may help sustain temperate environments.
}


   \maketitle
%
%-------------------------------------------------------------------

\section{Introduction}
\label{Intro}

The number of detected low-mass exoplanets has grown quickly over the past decades.
Nowadays, the most abundant planet classes are super-Earths and mini-Neptunes, that is, planets with masses larger than that of Earth and lower than that of Neptune \citep[e.g.][]{Fulton_etal_2017, Schlichting_2018}.
Because there is an observational bias towards large masses, a vast population of Earth-mass planets is also expected, as shown by formation studies \citep[e.g.][]{Schlecker_etal_2021}.
Many of these planets are currently found in the habitable zone \citep{Kasting_etal_1993} of low-mass M-dwarf and K-dwarf stars  \citep{Burn_etal_2021}. 
Therefore, we may wonder if they are similar to our Earth, and if so, which of these planets are possible candidates to harbour life as we know it.

When considering the habitability of a planet, not only does the host star have an influence but also the planetary properties, such as  the composition of the atmosphere and the spin dynamics \citep[e.g.][]{Kopparapu_2019}. 
The mass of super-Earths is not large enough to allow these planets to retain a voluminous gaseous envelope, so they likely have a rocky composition. 
A thin atmospheric layer as it is observed on Earth and Venus is nevertheless expected \citep[e.g.][]{Demory_etal_2016, Komacek_Abbot_2019, Wordsworth_Kreidberg_2022}, which can help sustain liquid water \citep{Kasting_etal_1993, Kopparapu_etal_2013}.
The spin of these bodies controls the heat distribution on the surface from the incoming stellar radiation, known as insolation \citep[e.g.][]{Milankovitch_1941,Ward_1974}, which strongly affects their surface temperature equilibrium and atmospheric general circulation \citep[e.g.][]{Forget_Leconte_2014, Dobrovolskis_2021}. 
Therefore, the spin is a key quantity in predicting planet's climate.

Low-mass stars have a lower luminosity than Sun-like stars.
As a result, the habitable zone is closer to the star, leading to significant star-planet interactions.
Some of these stars can be very active, such as M-dwarfs in particular \citep{Kopparapu_2019}, which can trigger significant atmospheric losses on close-in planets \citep[e.g.][]{Luger_Barnes_2015, Bolmont_etal_2017,France_etal_2020}.
Gravitational tides are also very strong, and they can modify the spin and the orbit of a planet until synchronous rotation is achieved \citep{Hut_1980, Adams_Bloch_2015}.
In this configuration, the rotation period is equal to the orbital period, implying that one side of the planet always faces the star and becomes extremely hot, while the other side becomes very cold. \footnote{This is the case in the limit of thin atmospheres. However, the day-night temperature difference depends on the efficiency of the heat redistribution mechanisms, which has been examined in the literature using various modelling approaches \citep[e.g.][]{Wordsworth_2015, Koll_Abbot_2016, Auclair-Desrotour_etal_2022}.}
Although synchronous rotation seems unfeasible for the development of life as we know it, it is not impossible \citep[e.g.][]{Yang_2013, Turbet_2016}.

The habitable zone of K-dwarf stars is further away, so planets in such areas are more protected from strong stellar winds.
However, the gravitational tidal interactions can still quickly modify their spins.
Planets with dense atmospheres also experience thermal tides, which result from the heating of the atmosphere by the star.
The torque induced by thermal atmospheric tides opposes that of gravitational tides, and thus it pushes the rotation away from the synchronisation.
Although the mass of the atmosphere is often negligible with respect to the mass of the mantle, thermal tides can be of the same order of magnitude \citep{Dobrovolskis_Ingersoll_1980}.
Indeed, the peculiar retrograde asynchronous rotation of Venus is believed to result from the balance between gravitational and thermal atmospheric tides \citep{Gold_Soter_1969, Ingersoll_Dobrovolskis_1978, Dobrovolskis_1980, Yoder_1997, Correia_Laskar_2001, Correia_Laskar_2003I, Auclair-Desrotour_etal_2017b, Revol_etal_2023}.

At present, thermal tides are negligible on Earth because they generate a torque of $\sim 3.2 \times 10^{15}$ J \citep[e.g.][]{Volland_1990}, that is, less than $10 \%$ of the gravitational (ocean $+$ solid) torque, a consequence of our planet being too distant from the Sun \citep{Correia_etal_2008}.
Nevertheless, for Earth-like planets in the habitable zone of K-dwarf stars, thermal atmospheric tides can lead to asynchronous spin-orbit equilibria similar to that of Venus \citep{Leconte_etal_2015}.
Moreover, thermal tides can also contribute to increasing the obliquity of these planets \citep[e.g.][]{Correia_Laskar_2003I, Cunha_etal_2015}.
Moderate obliquities give rise to more temperate climates because the obliquity controls the insolation distribution for a given latitude and is therefore the main driver of the seasons \citep[e.g.][]{Dobrovolskis_2009, Dobrovolskis_2021}.

In this paper we revisit the spin evolution of Earth-like planets around low-mass stars mainly in the range of K-dwarf stars.
For this purpose, we used a complete 3D dynamical model for the gravitational and thermal tides based on a vectorial formalism \citep{Correia_Valente_2022, Valente_Correia_2023}.
We adopted the Andrade rheological model to handle the tidal deformation of the rocky mantle \citep[e.g.][]{Efroimsky_2012a, Gevorgyan_etal_2020} and the output of general circulation models (GCMs) to simulate the thermal response of the atmosphere \citep[e.g.][]{Leconte_etal_2015, Auclair-Desrotour_etal_2019a}; the models are believed to provide realistic descriptions of the atmosphere dynamics of rocky planets.
Our aim is to better understand  the possible end states for the spin of these planets and their impact on habitability as a function of stellar type, distance to the star, and orbital eccentricity.

In Sect.~\ref{methods}, we present the model that we used to study the tidal evolution of an Earth-mass planet around a single star.
In Sect.~\ref{Venus}, we apply this model to Venus in order to gauge the unknown parameters, assuming that Venus is currently on an equilibrium state \citep{Correia_Laskar_2001}.
In Sect.~\ref{Earth}, we apply our model to an Earth-twin planet (same mass and atmosphere properties of the Earth) and subject it to the environment of a low-mass star.
We determine the equilibria for the spin for a wide range of stellar masses, semi-major axes, and eccentricities.
In Sect.~\ref{apliKepler}, we apply our model to the planet Kepler-1229\,b, a super-Earth with radius $1.4~R_\oplus$ in an 87-day orbit in the habitable zone of a star with mass $0.54~M_\odot$ \citep{Morton_etal_2016}, and compare with the general results from our model. 
Finally, in the last section, we summarise and discuss our results.

%%%%%%%%%%%%

\section{Model}
\label{methods}

We consider a system formed by a planet and a star moving in a Keplerian orbit, with masses $\mp$ and $\ms$, respectively.
The orbital angular momentum and orbital energy are respectively given by \citep[e.g.][]{Murray_Dermott_1999}
\be
\vG = \beta n a^2 \sq \, \vk 
\ ,\quad \mathrm{and} \quad
\Eo = - \frac{\Gc \ms \mp}{2 a} \ ,
 \llabel{eqS21}
\ee
where $\beta = \ms \mp (\ms + \mp)^{-1}$ is the reduced mass, $a$ is the semi-major axis, $e$ is the eccentricity, $n=[\Gc (\ms + \mp) a^{-3}]^{1/2}$ is the mean motion, $\Gc$ is the gravitational constant and $\vk$ is the unit vector along the direction of $\vG$, which is normal to the orbit.

The star is a point-mass object with luminosity $L_\osdot$, while the planet is an extended body composed by a mantle with radius $\Rp$, and surrounded by a thin atmosphere.
It rotates with angular velocity $\vw=\om \, \vs$, where $\vs$ is the unit vector along the direction of the spin axis. 
We assumed that $\vs$ is also the axis of maximal inertia (gyroscopic approximation), so the rotational angular momentum is simply given by
\be
\vL = C \vw \ , \llabel{eqS22}
\ee
where $C = \xi \mp \Rp^2$ is the principal moment of inertia of the planet and $\xi$ is an internal structure constant.
We further assumed that the mantle of the planet and its atmosphere can be deformed under the action of gravitational and thermal tides, respectively.

\subsection{Gravitational tides}
\llabel{sect:gtt}

Gravitational tides arise from differential and inelastic deformations of the mantle due to the gravitational perturbations of the star.
The atmosphere also undergoes gravitational tides, but they can be neglected. % \citep[e.g.][]{Auclair-Desrotour_etal_2017b}.
The distortion of the planet gives rise to a tidal potential \citep[e.g.][]{Lambeck_1980},
\begin{equation}
        V_g (\vr) = - \hat k_2 \frac{\Gc \ms}{\Rp} \left(\frac{\Rp}{r}\right)^3 \left(\frac{\Rp}{r_\osdot}\right)^3 P_2 ( \ur \cdot \ur_\osdot) \ , \llabel{eqS24}
    \end{equation}
where $\vr$ is the distance measured from the planet centre-of-mass, $r = ||\vr||$ is the norm, $\ur = \vr / r $ is the unit vector, $\vr_\osdot$ is the position of the star, $P_2(x) = (3 x^2 -1)/2 $ is a Legendre polynomial and $\hat k_2$ is the second Love number for potential.

The deformation of the mantle is characterised by the Love number, $\hat k_2$, which depends on the frequency of the perturbation, $\sigma$, and on the internal structure of the planet.
In general, we can write \citep[e.g.][]{Lambeck_1980}
    \begin{equation}
       \hat k_2 (\sigma) = \frac{\kf}{1 + \hat \mu (\sigma)} \ , \llabel{eqS219}
    \end{equation}
where $\kf$ is the fluid Love number and $\hat \mu$ is an effective rigidity. 
For a homogenous body, $\kf = 3/2$, but more generally it depends on the mass distribution inside the planet and can be related to the structure constant $\xi$ through the Darwin-Radau equation \citep[e.g.][]{Jeffreys_1976}. 
For Earth and Venus, we have $\kf =0.933$ and $\kf = 0.928$, respectively \citep{Yoder_1995cnt}. 

In order to compute the effective rigidity, $ \hat \mu (\sigma)$, we need to adopt some rheological model for the deformation.
The exact response of rocky planets to stress is unknown due to the lack of information about their interiors \citep[e.g.][]{Moquect_etal_2011}. %, Aitta_2012}.
However, viscoelastic rheologies are usually accepted as more realistic, because they are able to simultaneously reproduce the short-time and the long-time responses \citep[e.g.][]{Remus_etal_2012b, Efroimsky_2012b, Correia_etal_2014, Renaud_Henning_2018}.

Transient creep of many materials at high temperatures obeys the Andrade’s law \citep[e.g.][]{Louchet_Duval_2009}. 
Moreover, geophysical data from body and surface waves, free oscillations, and Chandler wobble, which cover a large spectrum of frequencies, also suggest that the mantle dissipates energy according to an Andrade rheology \citep[e.g.][]{Andrade_1910, Anderson_Minster_1979, Efroimsky_2012b, Gevorgyan_etal_2020}.
Therefore, we adopt here the Andrade viscoelastic model.
The effective rigidity is then given by \citep[e.g.][]{Efroimsky_2012b}
    \begin{equation}
        \hat \mu (\sigma) =  \frac{\tauv}{\taue} \left[1 - \ii (\sigma \taue)^{-1} + (\ii \sigma \ta)^{-\alpha} \Gamma (1+\alpha) \right]^{-1} \ , \llabel{eqS220}
    \end{equation}
where $\tauv$ is the viscous relaxation time, $\taue = \eta / \mu$ is the Maxwell relaxation time, $\eta$ is the effective viscosity and $\mu$ is the rigidity. 
The parameter $\alpha$ is the time exponent of the Andrade transient creep strain law, which is an empirical adjustable constant that usually ranges from $ 0.2 \le \alpha \le 0.4$ for rocky planets \citep{Castillo_Rogez_etal_2011}. 
The quantity $\ta$ is the timescale associated with the Andrade creep and may be termed as the ``Andrade'' or the ``anelastic'' time. 
In general, we have  $\ta \ge \taue$ \citep{Makarov_2012}, so for simplicity,\footnote{The impact of $\ta \ne \taue$ has been studied in \citet{Bolmont_etal_2020}.} we assumed that $\ta = \taue$.

The effective rigidity $\hat \mu$ is a complex number (Eq.\,\eqref{eqS220}), and hence also the Love number $\hat k_2$ (Eq.\,\eqref{eqS219}).
We can thus decompose $\hat k_2$ in its real and imaginary parts as
    \begin{equation}
        \hat k_2 (\sigma) = a_g (\sigma) - \ii \, \bg(\sigma) \ . \llabel{eqS25}
    \end{equation}
This partition is very useful because the imaginary part, $b_g (\sigma)$, is directly related to the amount of energy dissipated by tides.
It is therefore the main driver of the secular evolution of the system under the action of tides (see Sect.~\ref{sec:evol}).
For the Andrade model, we have from expression \eqref{eqS220}
    \begin{equation}
        \bg (\sigma) = \kf \, \frac{{\cal B} (\sigma) \, \sigma \tauv}{{\cal A} (\sigma)^2 + {\cal B} (\sigma)^2} = ( \kf - \ke ) \, \frac{{\cal B} (\sigma) \, \sigma \tau}{{\cal A} (\sigma)^2 + {\cal B} (\sigma)^2} \ , \llabel{eqS221} 
    \end{equation}
with
    \begin{equation}
        {\cal A} (\sigma) = (\sigma \tau) \left[ 1 + |\sigma \tau|^{-\alpha} \left(\frac{\taue}{\tau}\right)^{1-\alpha} \cos \left(\frac{\alpha \pi}{2}\right) \Gamma (1+\alpha) \right] \ , \llabel{eqS222}
    \end{equation}
and
    \begin{equation}
        {\cal B} (\sigma) = 1 + |\sigma \tau|^{1-\alpha} \left(\frac{\taue}{\tau}\right)^{1-\alpha} \sin \left(\frac{\alpha \pi}{2}\right) \Gamma (1+\alpha) \ , \llabel{eqS223}
    \end{equation}
where $\tau = \taue + \tauv$ is the total relaxation time and $\ke = \kf \, \taue/\tau$ is the elastic Love number.
For Earth and Venus, we have $\ke = 0.299$ and $ \ke = 0.25$, respectively \citep{Yoder_1995cnt}. %, which yields $\tau \approx 3 \taue$.

\subsection{Thermal atmospheric tides}

The differential absorption of the stellar heat by the planet’s atmosphere gives rise to local variations of temperature and consequently to pressure gradients. 
The mass of the atmosphere is then permanently redistributed, adjusting for the equilibrium. 
As for gravitational tides, the redistribution of mass in the atmosphere gives rise to an atmospheric bulge that modifies the gravitational potential generated by the atmosphere in any point of the space,
which is given by \citep[e.g.][]{Correia_etal_2003}
    \begin{equation}
        \VA = - \hat p_2 \frac{3}{5 \bar{\rho}}  \left( \frac{R}{r} \right)^3 \left( \frac{a}{r_\osdot} \right)^2 P_2 ( \ur \cdot \ur_\osdot) \ , \label{eqS224}
    \end{equation}
where $\bar{\rho}$ is the mean density of the planet and $\hat p_2$ is the second harmonic of the surface pressure variations, which depends on the frequency of the perturbation, $\sigma$, and on the composition and properties of the atmosphere.

In order to compute $\hat p_2 (\sigma) $, we need to adopt some dynamical model for the atmosphere.
For terrestrial planets in the habitable zone, the atmosphere can usually be treated as an ideal gas obeying the perfect gas law, together with the conservation of mass and the Navier-Stokes equations \citep[e.g.][]{Siebert_1961, Chapman_Lindzen_1970}. 
In addition, assuming a slowly rotating convective atmosphere, the Coriolis acceleration can be neglected and the hydrodynamic equations greatly simplified \citep[e.g.][]{Auclair-Desrotour_etal_2017a}.
Following \citet{Dobrovolskis_Ingersoll_1980}, thermal tides are considered to be generated solely by the average heat absorbed at the ground, $J_0$, which gives for the second harmonic of the surface pressure variations \citep[][Eq.\,(166)]{Auclair-Desrotour_etal_2017a}:
    \begin{equation}
        \hat p_2 (\sigma) = - \frac{\kappa \, \rho_0 \, J_0}{\sigma_0 + \ii \sigma} 
%        = -  \frac{\sqrt{\frac{10}{3\pi}} q_0}{1 + \ii (\sigma/\sigma_0)} 
        \ , \llabel{eqS232}
    \end{equation}
with $\kappa = 1- \gamma^{-1}$, where $\gamma$ is the adiabatic exponent value, $\rho_0$ is the mean surface density of the atmosphere, and $\sigma_0$ is the radiative cooling frequency of the atmosphere, which depends on its thermal capacity. 
As for the Love number (Eq.\,\eqref{eqS25}), we can also decompose $\hat p_2$ in its real and imaginary parts as
    \begin{equation}
        \hat p_2 (\sigma) = a_a (\sigma) - \ii \, b_a (\sigma) \ , \llabel{eqS225}
    \end{equation}
where the imaginary part characterises the viscous response of the atmosphere. 
The modulus gives the amplitude of the pressure variations and the argument gives the phase lag between the sub-stellar point and the maximal deformation of the atmosphere.
We thus have from expression \eqref{eqS232}
\begin{equation}
    \ba (\sigma) = -  \frac{ \kappa \, \rho_0 \, J_0 \, \sigma}{{\sigma_0}^2 + \sigma^2} 
    = -  \frac{\sigma/\sigma_0}{1 + (\sigma/\sigma_0)^2} \sqrt{\frac{10}{3\pi}} \, q_0
    \ , \llabel{eqS233}
\end{equation}
where $q_0$ is the amplitude of the quadrupole term of the pressure variations at zero frequency \citep{Leconte_etal_2015}.
This is the same functional form as in the model introduced by \citet{Ingersoll_Dobrovolskis_1978}.
The minus sign in the expression of $\ba (\sigma)$ causes the pressure variations to lead the star, a phenomenon well documented for Earth \citep[e.g.][]{Chapman_Lindzen_1970}. 
In this study we do not consider the possible emerging effects of high-frequency resonances, such as the Lamb waves resonance or the effect of varying the thermal forcing profile \citep[e.g.][]{Farhat_etal_2023, Laskar_etal_2023}.

\subsection{Secular evolution}
\label{sec:evol}

The additional disturbing potentials resulting from gravitational tides (Eq.\,\eqref{eqS24}) and thermal tides (Eq.\,\eqref{eqS224}) create a differential gravitational field around the planet given by
\be
\vg (\vr) = - \nabla_{\vr} \left[ V_g (\vr) + V_a (\vr) \right] \ .
\ee
The star itself interacts with this field; with mass $\ms$ and located at $\vr=\vr_\osdot$, it exerts a torque, $\vT$, on the planet that modifies its spin and orbit.
In an inertial frame we have 
\be
\dot \vG = \vT = \ms \, \vr_\osdot \times \vg  (\vr_\osdot)
\ , \llabel{150626a}
\ee
and, owing to the conservation of the total angular momentum,
\be
\dot \vL = - \dot \vG = - \vT
\ . \llabel{210805b}
\ee
The evolution of the orbital energy (power) is given by
\be
\dot \Eo = \ms \, \dot \vr_\osdot \cdot \vg (\vr_\osdot)
\ . \llabel{211110b}
\ee

In general, tidal effects slowly modify the spin and the orbit of the planet, on a timescale much longer than the orbital and precession periods of the system.
We can then average the torque (Eq.\,\eqref{150626a}) and the power (Eq.\,\eqref{211110b}) over the mean anomaly and the argument of the pericentre, and obtain the equations of motion for the secular evolution of the system.

Following \citet{Correia_Valente_2022},  we let ($\vp,\vq,\vs$) be a set of orthogonal vectors, such that
\be
\vp = \vk \times \vs  \quad \mathrm{and} \quad
\quad \vq = \vs \times \vp = \vk - x \, \vs  \ , 
\ee
with
\be
\label{230404a}
\cT = \cos \theta = \vk \cdot \vs \ ,
\ee
where $\vp$ is aligned with the line of nodes between the equator of the planet and the orbital plane, and $\theta$ is the angle between these two planes, also known as the obliquity.
For the average torque, we therefore have
\be
\big\langle \vT \big\rangle =  \left(T_p^g + T_p^a\right) \vp + \left(T_q^g + T_q^a\right) \vq + \left(T_s^g + T_s^a\right) \vs 
\ , \llabel{220405a}
\ee
and 
\be
\llabel{220405b}
\big\langle \dot \Eo \big\rangle = n \, (T_E^g + T_E^a)\ ,
\ee
where the superscript $^g$ refers to the contribution from gravitational tides (Sect.~\ref{torque:gt}) and the superscript $^a$ refers to contribution from thermal atmospheric tides (Sect.~\ref{torque:tam}).
The $T_p$ components in the torque (Eq.\,\eqref{220405a}) can be ignored because they change neither the norm of the angular momentum vectors nor the angle between them and therefore have no contribution to the secular spin and orbital evolution (see Sect.~\ref{GOSE}).

\subsubsection{Gravitational tides}
\llabel{torque:gt}

The complete expressions for the average torque and power due to gravitational tides, valid for any eccentricity, $e$, and obliquity, $\theta$, are given by \citet{Correia_Valente_2022},
    \begin{equation}
        \begin{split}
            \llabel{eqS212}
            T_q^g = \frac{3 \At_g}{32} \sum_{k=-\infty}^{+\infty} & \Bigg\{ 3 \, \bg ( -kn ) \left(1 - \cT^2 \right) \bigg[ \left( \HDPT \right)^2 - \left( \HDMT \right)^2 \bigg]  \\ 
            &  + 2 \, \bg ( \om-kn ) \, \bigg[  \left( 1+ \cT \right)^2 \left( 2-\cT \right) \left( \HDPT \right)^2  \\ 
            &  - 4 \cT^3 \left( \HZT \right)^2  - \left( 1- \cT \right)^2 \left( 2+\cT \right) \left( \HDMT  \right)^2  \bigg]  \\
            & + \bg \left( 2\om-kn \right) \, \bigg[ - 4 \cT \left( 1- \cT^2  \right) \left( \HZT \right)^2 \\
            &  + \left( 1+ \cT \right)^3 \left( \HDPT \right)^2 - \left( 1 - \cT \right)^3 \left( \HDMT \right)^2  \bigg] \Bigg\}  \ ,
        \end{split}
    \end{equation}
    \begin{equation}
        \llabel{eqS213}
        \begin{split}
            T_s^g = \frac{3 \At_g}{32} \sum_{k=-\infty}^{+\infty} & \Bigg\{ 2 \, \bg ( \om-kn ) \left( 1 - \cT^2 \right) \bigg[ 4 \cT^2 \left( \HZT  \right)^2   \\
            & + \left( 1 + \cT \right)^2 \left( \HDPT \right)^2 + \left( 1 -\cT \right)^2 \left( \HDMT  \right)^2  \bigg] \\
            & + \bg ( 2\om-kn ) \bigg[ 4 \left( 1 - \cT^2 \right)^2 \left( \HZT \right)^2  \\
            & + \left( 1 + \cT  \right)^4 \left( \HDPT  \right)^2 + \left( 1 - \cT  \right)^4 \left( \HDMT  \right)^2  \bigg] \Bigg\}   \ ,
        \end{split}
    \end{equation}
    \begin{equation}
        \llabel{eqS214}
        \begin{split}
            T_E^g = \frac{\At_g}{64}  \sum_{k=-\infty}^{+\infty} & k \, \Bigg\{ \bg ( -kn )  \bigg[
            4 \left(1-3 \cT^2\right)^2 \left(\HZT\right)^2 \\
            & + 9 \left(1-\cT^2\right)^2 \left(\left(\HDMT\right)^2 + \left(\HDPT\right)^2\right) \bigg]  \\
            & + 12 \, \bg ( \om-kn ) \left(1-\cT^2\right) \bigg[  4 \cT^2  \left(\HZT\right)^2 \\
            & +  \left(1-\cT\right)^2 \left(\HDMT\right)^2 + \left(1+\cT\right)^2 \left(\HDPT\right)^2  \bigg]  \\
            & + 3 \, \bg ( 2\om-kn )  \bigg[ 4 \left(1-\cT^2\right)^2 \left(\HZT\right)^2 \\
            & + \left(1-\cT\right)^4 \left(\HDMT\right)^2 + \left(1+\cT\right)^4 \left(\HDPT\right)^2 \bigg] \Bigg\} \ ,
        \end{split}
    \end{equation}
where
\be
\At_g = \frac{\Gc \ms^2 R^5}{a^6}  \ ,  \llabel{230405b}
\ee
and $X_k^{-\ell,m} (e)$ are the Hansen coefficients, which depend only on the eccentricity\footnote{In this work we consider only terms with $|k| \le 22$, which gives an error of  $e^{20} \sim 10^{-8}$ for the maximal adopted eccentricity ($e=0.4$).} (see Tab.~\ref{tabHansen}). 

\begin{table*}
\caption{Hansen coefficients up to $ e^6 $. \llabel{tabHansen} }
    \begin{center}
        \begin{tabular}{|r|c|c|c|c| } \hline 
            $k$ & $X_k^{-2,0} (e) $ & $X_k^{-2,2} (e) $ & $X_k^{-3,0} (e) $ & $X_k^{-3,2} (e) $ \\ \hline
            $-6$ & $\frac{1223}{320} e^6$ & $$ & $\frac{3167}{320} e^6$ & $$ \\
            $-5$ & $\frac{1097}{384} e^5$ & $$ & $\frac{1773}{256} e^5$ & $$ \\
            $-4$ & $\frac{103}{48} e^4 - \frac{129}{160} e^6$ & $-\frac{2}{45} e^6$ & $\frac{77}{16} e^4 + \frac{129}{160} e^6$ & $\frac{4}{45} e^6$ \\
            $-3$ & $\frac{13}{8} e^3 - \frac{25}{128} e^5$ & $-\frac{27}{640} e^5$ & $\frac{53}{16} e^3 + \frac{393}{256} e^5$ & $\frac{81}{1280} e^5$ \\
            $-2$ & $\frac{5}{4} e^2 + \frac{1}{6} e^4 + \frac{21}{64} e^6$ & $-\frac{1}{24} e^4 - \frac{3}{80} e^6$ & $\frac{9}{4} e^2 + \frac{7}{4} e^4 + \frac{141}{64} e^6$ & $\frac{1}{24} e^4 + \frac{7}{240} e^6$ \\
            $-1$ & $e + \frac{3}{8} e^3 + \frac{65}{192} e^5$ & $-\frac{1}{24} e^3 - \frac{13}{384} e^5$ & $\frac{3}{2} e + \frac{27}{16} e^3 + \frac{261}{128} e^5$ & $\frac{1}{48} e^3 + \frac{11}{768} e^5$ \\
            $0$ & $1 + \frac{1}{2} e^2 + \frac{3}{8} e^4 + \frac{5}{16} e^6$ & $$ & $1 + \frac{3}{2} e^2 + \frac{15}{8} e^4 + \frac{35}{16} e^6$ & $$ \\
            $1$ & $e + \frac{3}{8} e^3 + \frac{65}{192} e^5$ & $-e + \frac{3}{8} e^3 + \frac{7}{192} e^5$ & $\frac{3}{2} e + \frac{27}{16} e^3 + \frac{261}{128} e^5$ & $- \frac{1}{2} e + \frac{1}{16} e^3 - \frac{5}{384} e^5$ \\
            $2$ & $\frac{5}{4} e^2 + \frac{1}{6} e^4 + \frac{21}{64} e^6$ & $1 - \frac{7}{2} e^2 + \frac{29}{16} e^4 - \frac{53}{288} e^6$  & $\frac{9}{4} e^2 + \frac{7}{4} e^4 + \frac{141}{64} e^6$ & $1 - \frac{5}{2} e^2 + \frac{13}{16} e^4 - \frac{35}{288} e^6$ \\
            $3$ & $\frac{13}{8} e^3 - \frac{25}{128} e^5$ & $ 3 e - \frac{69}{8} e^3 + \frac{369}{64} e^5$ & $\frac{53}{16} e^3 + \frac{393}{256} e^5$ & $\frac{7}{2} e - \frac{123}{16} e^3 + \frac{489}{128} e^5$ \\
            $4$ & $\frac{103}{48} e^4 - \frac{129}{160} e^6$ & $\frac{13}{2} e^2 - \frac{55}{3} e^4 + \frac{239}{16} e^6$ & $\frac{77}{16} e^4 + \frac{129}{160} e^6$ & $\frac{17}{2} e^2 - \frac{115}{6} e^4 + \frac{601}{48} e^6$ \\
            $5$ & $\frac{1097}{384} e^5$ & $\frac{295}{24} e^3 - \frac{13745}{384} e^5$ & $\frac{1773}{256} e^5$ & $\frac{845}{48} e^3 - \frac{32525}{768} e^5$ \\
            $6$ & $\frac{1223}{320} e^6$ & $\frac{345}{16} e^4 - \frac{10569}{160} e^6$ & $\frac{3167}{320} e^6$ & $\frac{533}{16} e^4 - \frac{13827}{160} e^6$ \\
            $7$ & $$ & $\frac{69251}{1920} e^5$ & $$ & $\frac{228347}{3840} e^5$ \\
            $8$ & $$ & $\frac{42037}{720} e^6$ & $$ & $\frac{73369}{720} e^6$ \\ \hline
        \end{tabular} 
    \end{center}
    $X_k^{\ell,-2} (e) = X_{-k}^{\ell,2} (e)$. The exact expression of these coefficients is given by $  X_k^{-\ell,m} (e)  = \pi^{-1} \int_0^\pi \left( a/r \right)^{\ell} \exp(\ii m \lv) \exp(-\ii k M) \, d M $ .
\end{table*}

\subsubsection{Thermal atmospheric tides}
\llabel{torque:tam}

The complete expressions for the average torque and power due to thermal tides are given by \citet{Valente_Correia_2023},
    \begin{equation}
        \begin{split}
            \llabel{eqS229}
            T_q^a = \frac{3 \At_a}{32} \sum_{k=-\infty}^{+\infty} & \Bigg\{ 3 \, \ba ( -kn ) \, \left(1 - \cT^2 \right) \bigg[ \HDPT \HDPD - \HDMT \HDMD \bigg]  \\ 
            &  + 2 \, \ba ( \om-kn ) \, \bigg[  \left( 1+ \cT \right)^2 \left( 2-\cT \right) \HDPT \HDPD  \\ 
            &  - 4 \cT^3 \HZT \HZD - \left( 1- \cT \right)^2 \left( 2+\cT \right) \HDMT \HDMT  \bigg]  \\
            & + \ba \left( 2\om-kn \right) \, \bigg[ - 4 \cT \left( 1- \cT^2  \right)  \HZT \HZD \\
            &  + \left( 1+ \cT \right)^3 \HDPT \HDPD - \left( 1 - \cT \right)^3  \HDMT \HDMD  \bigg] \Bigg\}  \ ,
        \end{split}
    \end{equation}
    \begin{equation}
        \llabel{eqS230}
        \begin{split}
            T_s^a = \frac{3 \At_a}{32} \sum_{k=-\infty}^{+\infty} & \Bigg\{ 2 \, \ba ( \om-kn ) \left( 1 - \cT^2 \right) \bigg[ 4 \cT^2  \HZT \HZD   \\
            & + \left( 1 + \cT \right)^2  \HDPT \HDPD + \left( 1 -\cT \right)^2 \HDMT \HDMD \bigg] \\
            & + \ba ( 2\om-kn ) \bigg[ 4 \left( 1 - \cT^2 \right)^2 \HZT \HZD  \\
            & + \left( 1 + \cT  \right)^4  \HDPT \HDPD + \left( 1 - \cT  \right)^4  \HDMT \HDMD  \bigg] \Bigg\}   \ ,
        \end{split}
    \end{equation}
    \begin{equation}
        \llabel{eqS231}
        \begin{split}
            T_E^a = \frac{\At_a}{64}  \sum_{k=-\infty}^{+\infty} & k \, \Bigg\{ \ba ( -kn ) \, \bigg[ 4 \left(1-3 \cT^2\right)^2 \HZT \HZD \\
            & + 9 \left(1-\cT^2\right)^2 \left(\HDMT \HDMD + \HDPT \HDPD \right) \bigg]  \\
            & + 12 \, \ba ( \om-kn ) \, \left(1-\cT^2\right) \bigg[  4 \cT^2  \HZT \HZD \\
            & +  \left(1-\cT\right)^2 \HDMT \HDMD + \left(1+\cT\right)^2 \HDPT \HDPD \bigg]  \\
            & + 3 \, \ba ( 2\om-kn ) \, \bigg[ 4 \left(1-\cT^2\right)^2  \HZT \HZD \\
            & + \left(1-\cT\right)^4  \HDMT \HDMD + \left(1+\cT\right)^4  \HDPT \HDPD \bigg] \Bigg\} \ ,
        \end{split}
    \end{equation}
where 
\be
 \At_a = \frac{3 \ms}{5 \bar{\rho}} \left( \frac{R}{a} \right)^3 \ . \llabel{230405c}
 \ee

\subsection{Spin and orbital evolution} 
\label{GOSE}

The set of equations (\ref{220405a}) and (\ref{220405b}) allows us to track the secular evolution of the system using the angular momentum vectors and the orbital energy.
Concerning the spin, the rotation rate is directly given from the rotational angular momentum (Eq.\,(\ref{eqS22})),
\be
\om = \frac{\vL \cdot \vs}{C} = \frac{\sqrt{\vL \cdot \vL}}{C}
\ , \llabel{211027c}
\ee
while the angle between the orbital and equatorial planes (also known as obliquity) is obtained from both angular momentum vectors as (Eq.\,\eqref{230404a})
\be
\cos \theta = \vk \cdot \vs = \frac{\vG \cdot \vL}{\sqrt{(\vG \cdot \vG) (\vL \cdot \vL)}} 
\ . \llabel{211027z}
\ee
Concerning the orbit, the semi-major axis is directly given from the orbital energy (Eq.\,(\ref{eqS21}))
\be
a = - \frac{\Gc \ms \mp}{2 \Eo}
\ , \llabel{211110f}
\ee
while the eccentricity is obtained from the orbital angular momentum (Eq.\,(\ref{eqS21}))
\be
e  = \sqrt{1 - \frac{(\vG \cdot \vk)^2}{\beta^2 n^2 a^4}}
= \sqrt{1 - \frac{\vG \cdot \vG}{\beta^2 n^2 a^4}}
\ . \llabel{211110g}
\ee

%%%%%%%%%%

\section{Venus}
\label{Venus}

Venus has a peculiar spin state among all terrestrial planets \citep[e.g.][]{Smith_1963,Carpenter_1964};
it has a slow retrograde rotation, with a rotation period of $243.023 $~day ($\om / n = 0.92462$) and an obliquity $\theta = 177.361^\circ$ \citep{Margot_etal_2021}.
The present rotation of Venus may represent an equilibrium between gravitational and thermal atmospheric tides \citep{Gold_Soter_1969, Correia_Laskar_2001, Auclair-Desrotour_etal_2017a}. 
It is thus a unique example in our Solar System of the tidal mechanisms described in Sect.~\ref{methods}, which can be used to benchmark our model.

\subsection{Physical properties}
\llabel{ssct:pp}

The mass, radius, internal structure constant and Love numbers of Venus are well know (Tab.~\ref{tab:venus}).
The properties and composition of its atmosphere are also relatively well determined.
Therefore, it is possible to use general circulation models to compute the amplitude of the torque induced by thermal tides \citep[e.g.][]{ Leconte_etal_2013, Leconte_etal_2015, Auclair-Desrotour_etal_2019a, Wu_etal_2023}.
Indeed, in the case of Venus it has been shown that Eq.\,\eqref{eqS233} complies with the thermal response of the atmosphere at various frequencies and it was estimated that $\sigma_0 = 23.8$~yr$^{-1}$ and $q_0 = 2.01 $~mbar (Tab.~\ref{tab:venus}). %\citep{Leconte_etal_2015}. % $\sigma_0 = 2 \om_0$

\begin{table}
\caption{Physical and dynamical parameters adopted for the planets.  \label{tab:venus} } 
\begin{center}
{ \begin{tabular}{cccc} \hline \hline
Parameter & Venus & Earth & Kepler-1229\,b  \\ \hline
%age & $\sim 5$~Gyr \\ 
$\ms$ [$M_\odot$] & $1.00$ & $1.00$ & $0.54$ \\
$\mp$ [$M_\oplus$]  & $0.82$ & $1.00$ & $2.74$ \\
$\Rp$ [$R_\oplus$] & $0.95$ & $1.00$ & $1.40$ \\
$a$ [au] & $0.723$ & $1.000$ & $0.313$ \\ \hline
$F_\osdot$ [Wm$^{-2}$] & $2610$ & $1366$ & $606$ \\
$p_s$ [bar] & $92.$ & $1.$ & $10.$\\
$q_0$ [mbar] & $2.01$ & $11.8$ & $26.9$ \\
$\sigma_0$ [yr$^{-1}$]& $23.8$ & $145.$ & $85.3$ \\ \hline
$\xi$ & $0.330$ & $0.331$ & $0.331$ \\
$\ke$ & $0.25$ & $0.299$ & $0.299$\\
$\kf$ & $0.928$ & $0.933$ & $0.933$\\ 
$\alpha$ & $0.3$ & $0.3$ & $0.3$\\ 
$\taue$ [yr] & $1468$ & $1468$ & $1468$ \\ \hline
\end{tabular}}
\end{center}
The atmosphere parameters for Earth and Venus are from \citet{Leconte_etal_2015}, and for Kepler-1229\,b they are from \citet{Auclair-Desrotour_etal_2019a}.
The inner structure parameters such as the internal structure constant, $\xi$; the elastic Love number, $\ke$; and the fluid Love number, $\kf$, are from \citet{Yoder_1995cnt}, and for Kepler-1229\,b they are made equal to the Earth's values.
The remaining parameters of the Kepler-1229 system are from \citet{Morton_etal_2016}, where  $\mp$ is estimated from $\Rp$, assuming an Earth-like density. 
\end{table}

%
%    \begin{equation}
%        \ba (\sigma) = - \sqrt{\frac{10}{3 \pi}} \frac{ 2 q_0 \, \om_0 \, \sigma}{(2 \om_0)^2 + \sigma ^2} 
%        \ . \llabel{eqS234}
%    \end{equation}
%    
%
%    \begin{equation}
%        q_0 = \sqrt{ \frac{3 \pi}{10}} \frac{ \kappa \rho_0  J_0}{\sigma_0} \  , \quad \om_0 = \frac{\sigma_0}{2}. 
%    \llabel{eqS235}
%    \end{equation}
%
 
However, the rigidity and viscosity of Venus' mantle are not much constrained.
Since Venus is similar to Earth in mass and size, we can adopt the Earth's parameters.
Indeed, the rigidity and viscosity obtained for Venus using internal structure models with Andrade rheology are consistent with the values estimated for the Earth \citep{Bolmont_etal_2020, Melini_etal_2022, Saliby_etal_2023}.
The rigidity of the Earth's upper mantle is $\mu \approx 80$~GPa \citep{Karato_Wu_1993}, but the whole mantle rigidity can be larger \citep[see Table~2.1 in][]{Sabadini_etal_2016}.
By volume-averaging the rigidity value we get $\mu \approx 176$~GPa.
The average effective viscosity is much more uncertain, $\eta \sim 10^{21}$~Pa\,s \citep{Karato_Wu_1993}, which gives for the Maxwell relaxation time $\taue \sim 200$~yr \citep[for more details see][]{Bolmont_etal_2020}. % $\taue = \eta / \mu \sim 400 - 180$, respectively.
Actually, the viscosity may vary by several orders of magnitude, depending on the depth and temperature \citep[e.g.][]{Kirby_Kronenberg_1987}.
In the case of Earth, the surface post-glacial rebound due to the last glaciation about $10^4$~yr ago is still ongoing, suggesting that the Earth's mantle relaxation time can be $\taue \sim 4000$~yr \citep{Turcotte_Schubert_2002}.

\subsection{Dynamical constraints on Venus' mantle}
\llabel{ssct:dc}

The evolution of the rotation rate can be obtained from expressions \eqref{220405a} and \eqref{211027c} as
\be
\dot \om = - \frac{1}{C}\, \big\langle \vT \big\rangle  \cdot \vs = - \frac{T_s^g + T_s^a }{C} \, 
\ . \llabel{220407a}
\ee
The current eccentricity of Venus ($ e = 0.007 $) and its spin state\footnote{The current spin state of Venus can be described either by the pair $(\omega, \theta)$ or by the pair  $(-\omega, \pi-\theta)$. These two points do not correspond to the same physical state, but they are equivalent from a dynamical point of view \citep{Correia_Laskar_2001}.} ($\om_e / n = - 0.92462$, $\theta_e=2.639^\circ$),  are well determined \citep{Margot_etal_2021}, where $\om_e$ and $\theta_e$ are the current equilibrium rotation and obliquity, respectively. 
For simplicity, we can consider that Venus evolves in a circular orbit ($e \approx 0$) with zero obliquity ($\theta \approx 0$), and expression \eqref{220407a} simplifies as (Eqs.\,\eqref{eqS213} and \eqref{eqS230})
\begin{equation}
\dot \om = -  \frac{3 \At_g}{2 C} \bg \left( 2 \om -2n \right) - \frac{3 \At_a}{2 C} \ba \left( 2 \om -2n \right) 
\ . \label{230405a}
\end{equation}

Assuming that the present rotation of Venus is in an equilibrium state,\footnote{Because Venus is close to synchronous rotation, it is often assumed that it already reached an equilibrium state, although this hypothesis is yet to be confirmed \cite[e.g.][]{Margot_etal_2021, Revol_etal_2023}.} we must have $\dot \om = 0$ for $\om = \om_e$, and thus
%\be
%T_s^g = - T_s^a \ . \llabel{230406a}
%\ee
%
\begin{equation}
    \At_g \bg \left( 2 \om_e -2n \right) = - \At_a \ba \left( 2 \om_e -2n \right) \ . \label{eqS343}
\end{equation}
The constant parameters $\At_g$ (Eq.\,\eqref{230405b}) and $\At_a$ (Eq.\,\eqref{230405c}), along with the surface pressure variation function, $b_a \left( 2 \om_e -2n \right)$ (Eq.\,\eqref{eqS233}), are well determined for Venus (Tab.~\ref{tab:venus}).
The major source of uncertainty is related to the Love number function, $b_g \left( 2 \om_e -2n \right)$, which depends on the mantle rheology.
For the Andrade model (Sect.~\ref{sect:gtt}), this uncertainty is transferred to the determination of the $\alpha$ and $\taue$ parameters (Eq.\,\eqref{eqS221}).

For materials deforming according to an Andrade rheology with linear work hardening, we have $\alpha = 1/3$ \citep{Louchet_Duval_2009}.
However, for rocky planets it is commonly accepted that $0.2 < \alpha < 0.4$  \citep{Castillo_Rogez_etal_2011}.
Indeed, for Earth, \citet{Tobie_etal_2019} have shown that $\alpha \in \left[0.23, 0.28 \right]$ is able to correctly reproduce the observed tidal dissipation rate.
\citet{Melini_etal_2022} have studied the tidal deformation of Venus with the \texttt{ALMA$^3$} code (plAnetary Love nuMbers cAlculator), %, version 3). 
and conclude that $\alpha = 0.3$ provides the best agreement with tidal constraints on the interior of Venus determined by \citet{Dumoulin_etal_2017}.

\begin{figure}
\centering
\includegraphics[width=\columnwidth]{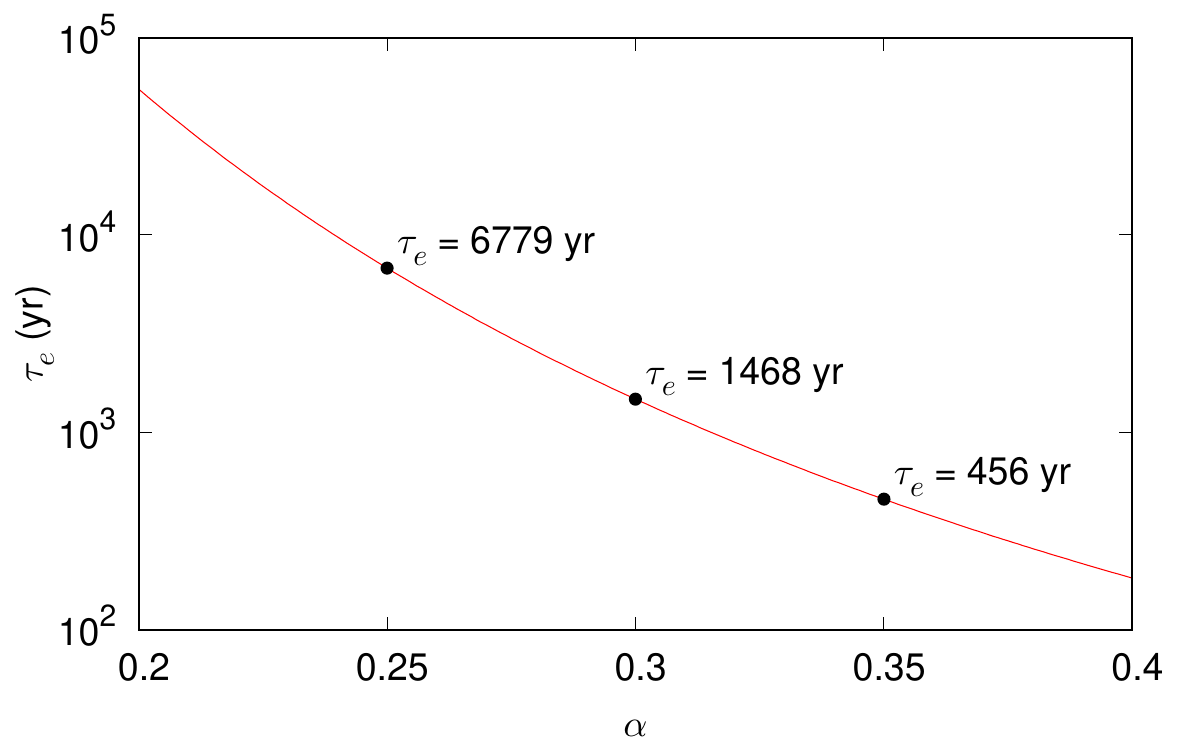}
\caption{Maxwell relaxation time $\taue$ as a function of the $\alpha$ parameter for Venus (Eq.\eqref{eqS343}) obtained for the present rotation state.} \label{Fig:3.1}
\end{figure}

In the case of Venus, we can use the dynamical constraint provided by the present spin state to find a link between $\alpha$ and $\taue$. 
In Fig.~\ref{Fig:3.1}, we show the solution of Eq.\,\eqref{eqS343} for these two parameters.
We observe that $\taue$ can vary by several orders of magnitude for the expected range of $\alpha \in \left[0.2, 0.4 \right]$.
For $\alpha = 0.2$ it is about 60~kyr, while for $\alpha = 0.4$ it is only about 200~yr.
Following \citet{Melini_etal_2022}, we adopt $\alpha = 0.3$, which gives for the Maxwell relaxation time, $\taue = 1468$~yr.
These values are consistent with those of the Earth (Sect.~\ref{ssct:pp}), so we adopted them for this class of planets in the remainder of this work (see Tab.~\ref{tab:venus}).

\subsection{Equilibrium states}
\llabel{ssct:es}

Using the tidal models for Venus' mantle and atmosphere that better match the current observations (Sect.~\ref{ssct:pp} and \ref{ssct:dc}), we can quickly investigate all the possible equilibria for its spin.

\begin{figure}
\centering
\includegraphics[width=\columnwidth]{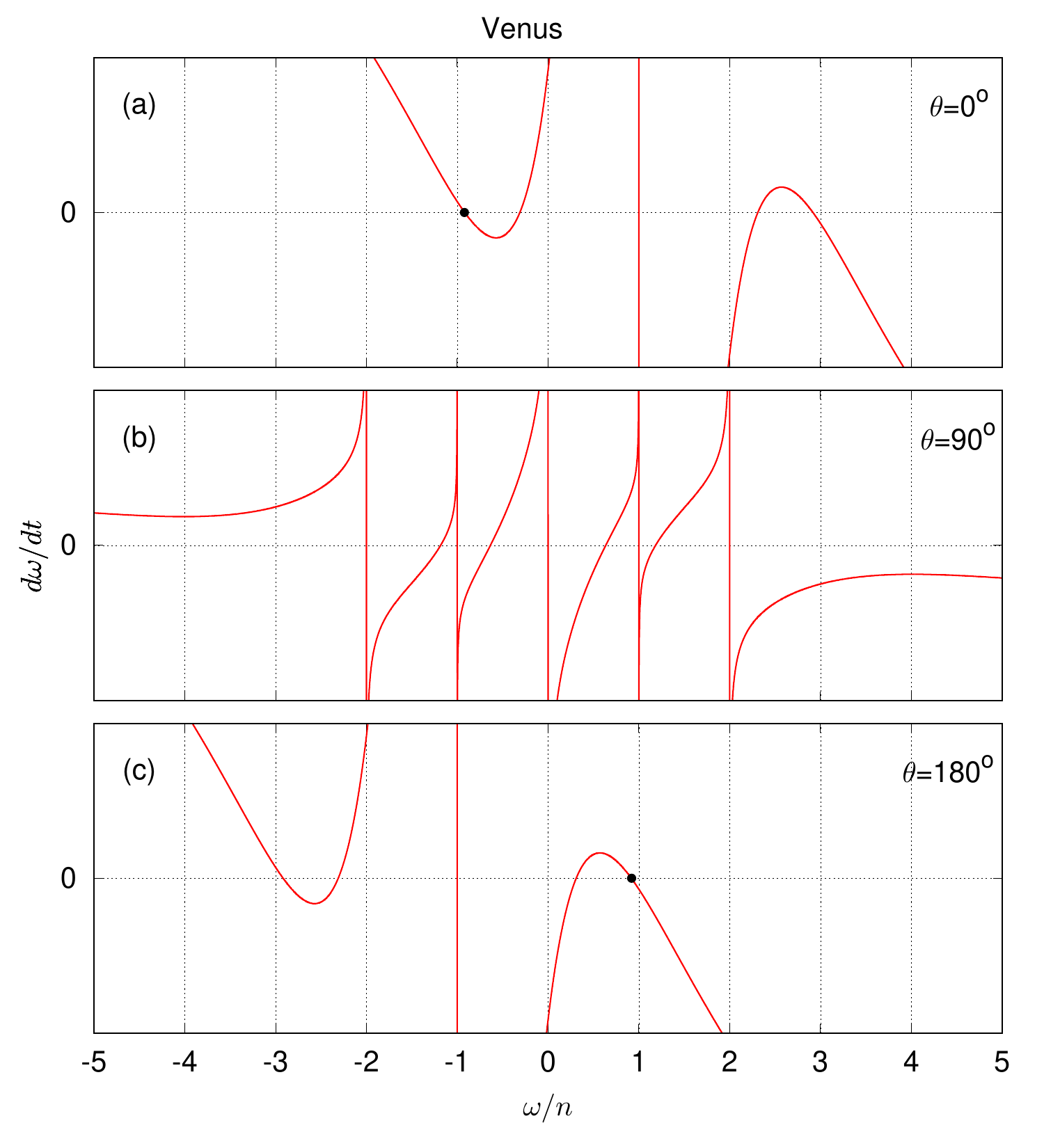}
\caption{Variation of $d \om/ dt $ upon $\om / n$ for Venus with $e=0$ (Eq.\,\eqref{220407a}). We show the evolution for different obliquity values: $\theta = 0^\circ$ (top), $\theta = 90^\circ$ (middle), and $\theta = 180^\circ$ (bottom plot). The black dots represent the presently observed rotation state.} \label{Fig:3.2}
\end{figure}

In Fig.~\ref{Fig:3.2}\,a, we show the evolution of the rotation rate for zero obliquity (Eq.\,\eqref{230405a}).
We observe that there are five equilibrium points ($\dot \om = 0$), though only three correspond to the stable equilibria (those with $\partial \dot \om / \partial \om < 0$).
One of these points corresponds to $\om = n$ (synchronous rotation), while the remaining two correspond to the present state and its symmetric, $\om - n = \pm \om_s$, with $\om_s/n = 1.92$ \citep{Correia_Laskar_2001}.
The current observed state, $\om_e = n - \om_s$, is marked with a dot.
The unstable equilibria are located at $\om - n = \pm \om_c$, with $\om_c/n = 1.31$. 
These unstable points are also important to understand the behaviour of the rotation rate.
For $\om/n \in [-0.31, 2.31]$, the rotation will evolve into the synchronous state, while outside this interval it will move to the closest asynchronous stable state, $\om = n \pm \om_s$.
These results are in agreement with previous studies that used different tidal models \citep{Correia_Laskar_2001, Correia_Laskar_2003I, Leconte_etal_2015, Auclair-Desrotour_etal_2017b}.

Because our model is valid for any obliquity value (Sect.~\ref{methods}), we can study the evolution of the rotation rate at any obliquity (Eq.\,\eqref{220407a}).
In Fig.~\ref{Fig:3.2}\,c, we show the evolution for $\theta = 180^\circ$. 
As expected, this case is similar to the case with $\theta = 0^\circ$, since the pair $(\omega, 180^\circ)$ behaves like the pair $(-\omega, 0^\circ)$ \citep{Correia_Laskar_2001}.
However, assuming that Venus rotation rate brakes from initial fast rotations, we always come from the right-hand side of the figure.
As a consequence, for $\theta=0^\circ$, Venus is expected to evolve into the state $\om=n+\om_s$, while for $\theta=180^\circ$, it is expected to end up in the state $\om=n-\om_s$ (present state).

The previous two obliquity values ($\theta = 0^\circ$ and $180^\circ$) are the most likely outcomes of gravitational and thermal tides \citep{Correia_Laskar_2001, Correia_Laskar_2003I, Correia_etal_2003}.
However, we cannot rule out that other possibilities for the obliquity exist, at least temporarily.
Therefore, in Fig.~\ref{Fig:3.2}\,b, we also show the evolution for $\theta = 90^\circ$.
Interestingly, we observe that the asynchronous thermal states are no longer possible.
On the other hand, the spin evolution is dominated by spin-orbit resonances that were absent before ($\om / n = -2,-1,0,+1,+2$).
These resonant states result from the geometry of the tidal torques (Eqs.\,\eqref{eqS221} and \eqref{eqS233}), which experience a significant increment at these specific rotation rates (Eqs.\,\eqref{eqS213} and \eqref{eqS230}).
As a result, the spin will settle into one of these states and stay there until the obliquity evolves into more modest values.
Although $\theta=90^\circ$ may not correspond to a final possibility for the spin of Venus, the crossing of the spin-orbit resonances can modify its final evolution (see next section).

\subsection{Complete spin evolution}
\llabel{ssct:cse}

The rotation rate and the obliquity evolution cannot be dissociated (Eqs.\,\eqref{211027c} and \eqref{211027z}).
Therefore, the exact evolution of the spin can only be obtained by integrating Eq.\,\eqref{210805b}.
This allows us to check how Venus may have evolved into the different equilibrium states (Sect.~\ref{ssct:es}) and also the existence of some eventual nonzero obliquity equilibria.

The orbital evolution (Eqs.\,\eqref{150626a} and \eqref{211110b}) is negligible in the case of Venus because $||\vL|| \ll ||\vG||$.
Actually, the eccentricity of Venus undergoes much stronger variations due to planetary perturbations than due to tidal effects \citep{Correia_Laskar_2003I}.
The present eccentricity of Venus is about 0.007 and its mean value is approximately 0.022 \citep{Laskar_2008}.
For simplicity, we thus consider that Venus evolves in a circular orbit ($e \approx 0$) and that $\vG$ and $\Eo$ are constant (Eq.\,\eqref{eqS21}).

\begin{figure}
\centering
\includegraphics[width=\columnwidth]{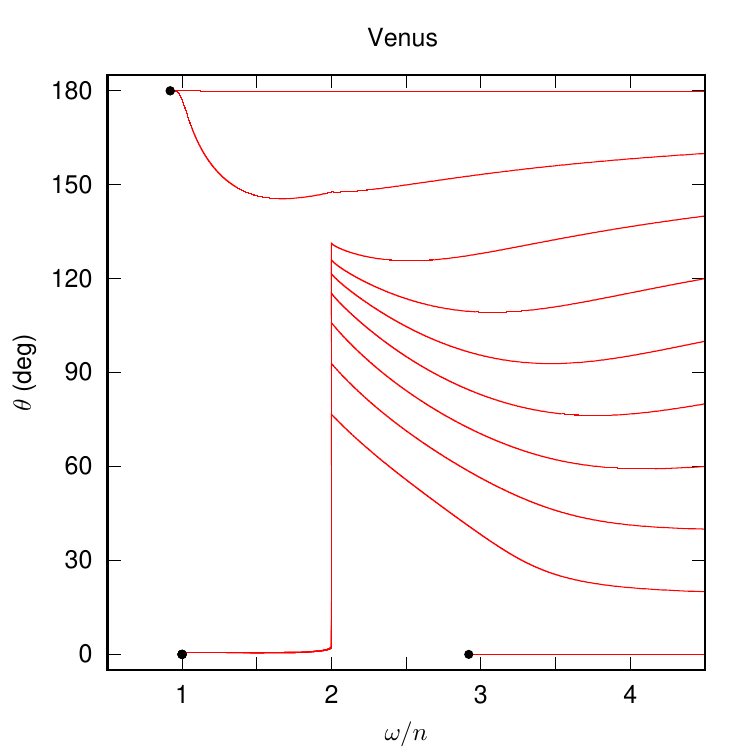}
\caption{Secular evolution of the spin of Venus in the plane ($\omega/n, \theta$) for different initial obliquity values. This plot was obtained with an initial $\om/n = 4.5$ and by integrating Eq.\,\eqref{210805b}, where the Hansen coefficients $X_k^{\ell,m} (e)$ are truncated at $|k|\leq 22$. 
The stable equilibrium points are marked with a black dot.} \label{Fig:3.3}
\end{figure}

In Fig.~\ref{Fig:3.3}, we show some trajectories for the tidal evolution of the spin in the plane $(\omega/n, \theta)$.
We start the integrations with $\om /n = 4.5 $, % which corresponds to a rotation period of $P \sim 50$~day,
and different initial obliquity values in the range $\left[ 0^\circ, 180^\circ \right]$.
Tidal effects initially decrease the rotation rate (Eq.\,\eqref{220407a}), so the evolutionary paths must be followed for decreasing values of $\om/n$, until they reach an equilibrium position (marked with a black dot).

For the initial low obliquity ($\theta = 0^\circ$), the obliquity does not change much and the rotation rate reaches the asynchronous rotation with $\om / n = 2.92$, corresponding to the prograde final state ($\om = n + \om_s, \theta = 0^\circ$).
For initial high obliquity ($\theta \ge 160^\circ$), the obliquity evolves into $180^\circ$ and then the rotation rate reaches the asynchronous rotation with $\om / n = 0.92$, which is equivalent to the present retrograde final state ($\om = n - \om_s, \theta = 0^\circ$).

For initial obliquities ranging from $20^{\circ}$ to $140^{\circ}$, the rotation rate can be locked in the 2/1 spin-orbit (see also Fig.~\ref{Fig:3.2}\,b).
At that point, the obliquity decreases until it reaches a much smaller value.
At the critical value $\theta \approx 3^\circ$, the 2/1 resonance becomes unstable and the rotation rate is allowed to decrease again.
It then moves to the synchronous rotation, corresponding to the final state ($\om = n, \theta = 0^\circ$).
This behaviour can be perfectly understood by looking at Fig.~\ref{Fig:3.2}\,a.
For near zero obliquities, rotation rates larger than the critical value $\om/n = 2.31$ move into the asynchronous final state $\om = n + \om_s$, while rotation rates smaller than these critical value move into the synchronous state.

Our results are in perfect agreement with those obtained by \citet{Revol_etal_2023}, who studied the spin evolution of Venus-like planets using different values for the $\alpha$ and $\taue$ parameters.
However, our results do not explain so well the current observations, since the present state can only be reached by a small set of initially extremely high obliquities ($\theta \ge 160^\circ$).
This is likely because in the present study we are not including all the effects that contribute to modify the spin evolution of Venus. 
Indeed, we do not take into account either the effect of core-mantle friction, which tends to align the equator of the planet with the orbital plane \citep{Correia_etal_2003}, or the effect of planetary perturbations, which can introduce chaotic variations on the obliquity \citep{Correia_Laskar_2001, Correia_Laskar_2003I}.

%%%%%%%%%%

\section{Earth-like planets}
\label{Earth}
        
A configuration similar to the Earth around the Sun is ideal when searching for habitable worlds.
However, low-mass stars are more frequent than sun-like stars in the Milky Way \citep[e.g.][]{Bochanski_2010}. 
The occurrence rate of planets around low-mass stars is also higher because they are easier to detect \citep[e.g.][]{Winn_2018}.
Moreover, formation studies show that temperate Earth-sized planets may be more frequent around M and K-dwarf stars \citep{Burn_etal_2021}.
The fact that Earth-like planets can be relatively common around low-mass stars offers a great opportunity to explore the possibility of having life.

The mean surface pressure on Earth is only $p_s = 1$~bar, which contrasts with that of Venus, $p_s = 92$~bar (Tab.~\ref{tab:venus}).
However, the amplitude of the thermal tide for an Earth-like atmosphere is about an order of magnitude larger, because on Venus the sunlight is almost completely scattered or absorbed before it reaches the surface \citep{Leconte_etal_2015}. 
As a result, Earth-like planets around low-mass stars can develop non-synchronous rotation similar to that of Venus \citep{Correia_etal_2008, Cunha_etal_2015, Leconte_etal_2015}.

In this section, we revisit the equilibria for the spin of Earth-like planets around low-mass stars.
We benefited from more accurate models for both gravitational and thermal atmospheric tides (Sect.~\ref{methods}), which allows us to perform a more comprehensive study of all the possibilities for a wide range of stellar masses, semi-major axes, and eccentricities.
We first explain how we obtain the stellar mass as a function of the semi-major axis that preserves the average flux received by the planet (Sect.~\ref{ss:smass}).
We then determine the number of possible equilibrium states for a given semi-major axis (Sect.~\ref{efs:earth}) and show how the position of the thermal asynchronous non-resonant equilibria change with the semi-major axis and with the eccentricity (Sect.~\ref{ss:anrts}).
Finally, we generalize these results for any planet in the habitable zone, that is, with an arbitrarily average stellar flux (Sect.~\ref{HZ}).

\subsection{Physical properties}

We consider an Earth-twin planet around a low-mass star, that is, we assume the exact same mass, radius, atmosphere composition and average annual stellar flux as the Earth's
(Tab.~\ref{tab:venus}).
It is then possible to run general circulation models to compute the amplitude of the torque induced by thermal tides, which was done in \citet{Leconte_etal_2015}.
For an average flux $F_\osdot =1366$~Wm$^{-2}$, they obtain\footnote{The response of the atmosphere in \citet{Leconte_etal_2015} is obtained using general circulation models for a planet around a Solar-type star. We note that, eventually, the same average stellar flux generated by a different stellar type star may not be absorbed exactly in the same way.} for the radiative frequency of the atmosphere, $\sigma_0 = 145$~yr$^{-1}$, % $\sigma_0 = 2 \om_0$ 
and for the amplitude of the atmospheric quadrupole, $q_0 = 11.8$~mbar.
For the gravitational tides we adopt the Andrade model with the parameters $\alpha=0.3$ and $\taue = 1468$~yr (Sect.~\ref{ssct:dc}).

\subsection{Stellar mass}
\label{ss:smass}

As we fix the average incoming stellar flux, we can determine the stellar luminosity from the orbital parameters of the planet through the following relation \citep[e.g.][]{Laskar_etal_1993AA}
\begin{equation}
F_\osdot = \left\langle \frac{L_\osdot}{4 \pi r_\osdot^2} \right\rangle = \frac{L_\osdot}{4 \pi a^2 \sqrt{1-e^2}}
  \ . \label{eqS446}
\end{equation}
We then use the mass-luminosity relation to obtain the stellar mass. 
For the low-mass stars under investigation, this relation is obtained for $0.20 < \ms /  M_\odot < 0.85 $ as \citep{Cuntz_Wang_2018},
\begin{equation}
    \frac{L_\osdot}{L_{\odot}} \approx \left( \frac{\ms}{M_{\odot}} \right)^{f(\ms/M_\odot)} \label{eqS444},
\end{equation}
with
\begin{equation}
    f(x) = -141.7 x^{4} + 232.4 x^{3} -129.1 x^{2} +33.29 x+0.215 \ , \label{eqS445}
\end{equation}
where $L_{\odot}$ and $M_\odot$ are the Sun luminosity and mass, respectively. 
By solving Eqs.~\eqref{eqS446} and \eqref{eqS444}, we get the stellar mass as a function of the semi-major axis and eccentricity for a constant flux.

\subsection{Equilibrium final states}
\llabel{efs:earth}

\begin{figure}
\centering
\includegraphics[width=\columnwidth]{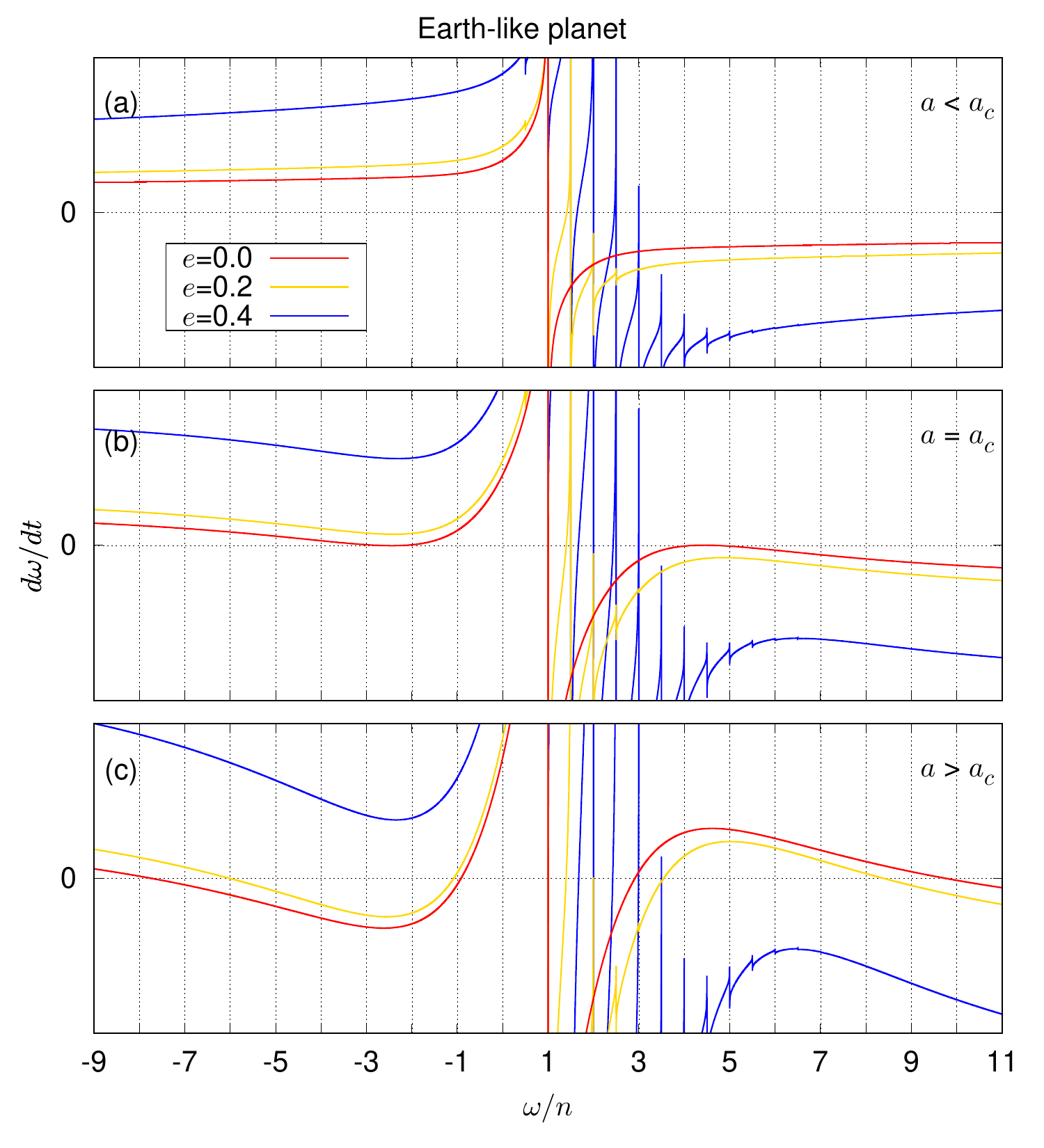}
\caption{Variation of $d \om/ dt $ upon $\om / n$ for an Earth-like planet with $\theta = 0^\circ$  (Eq.\,\eqref{eqS447}). We show the evolution for different semi-major axes: $a / a_c =0.63$ ($\ms / M_\odot \approx 0.52$) (top), $a=a_c=0.32$~au ($\ms / M_\odot \approx 0.68$) (middle), and $a/a_c =1.10$ ($\ms / M_\odot \approx 0.70$) (bottom plot). We also show the evolution for three different values of the eccentricity: $e=0.0$ (red line), $e=0.2$ (orange line),  and $e=0.4$ (blue line).} \label{Fig:4.1}
\end{figure}

In the absence of thermal atmospheric tides, %and for circular orbits, 
the torque resulting from gravitational tides (Sect.~\ref{torque:gt}) drives the spin into synchronous rotation, $\om / n = 1$, and zero obliquity, $\theta = 0^{\circ}$ \citep[e.g.][]{Hut_1980}. 
In the presence of thermal tides, an additional torque on the spin emerges (Sect.~\ref{torque:tam}) that can counter the braking effect from gravitational tides and give rise to additional equilibrium possibilities.
In order to investigate the multitude of scenarios, in Fig.~\ref{Fig:4.1}, we show the evolution of the rotation rate with zero obliquity for different values of the semi-major axis and eccentricity (Eq.\,\eqref{220407a}),
\begin{equation}
\begin{split}
     \dot \om =\frac{3}{2 C} \sum_{k=-\infty}^{+ \infty} & \bigg[ \At_g \bg \left( 2 \om -kn \right) \left( X_k^{-3,2} \right)^2 \\ & +\At_a  \ba \left( 2 \om -kn \right) X_k^{-3,2} X_k^{-2,2} \bigg] \ . \label{eqS447}
\end{split}
\end{equation}
Equilibrium points are observed for $\dot \om = 0$, and they can be stable for negative slopes ($\partial \dot \om / \partial \om < 0$). 

For circular orbits (Fig.~\ref{Fig:4.1}, red curves), there is a critical semi-major axis ($a_c \approx 0.32$~au) bellow which the spin behaves as in absence of thermal tides, that is, there is only one equilibrium possibility, corresponding to the synchronous rotation (Fig.~\ref{Fig:4.1}\,a).
For an Earth-like planet, this critical threshold corresponds to a stellar mass $\ms / M_\odot \approx 0.68$.
However, for larger semi-major axes (hence, for larger stellar masses), we observe that there are five equilibrium points (Fig.~\ref{Fig:4.1}\,c), as in the case of Venus (Sect.~\ref{ssct:es}).
Again, three of these points can be stable, one for $\om = n$ (synchronous rotation) and the other two for $\om = n \pm \om_s$.
Here, $\omega_s = |\om - n|$ is the synodic rotation frequency (that is, the frequency between two passages of the star in the local meridian), which depends on the balance between thermal and gravitational tides.
We thus confirm that the rotation of Earth-like planets in circular orbits around low-mass stars can also be locked in asynchronous states \citep{Leconte_etal_2015}.
The emergence of these new equilibria is favoured for planets further away from the star, because the relative strength between thermal and gravitational tides increases for larger semi-major axis \citep{Correia_etal_2008}.
Indeed, from expressions \eqref{230405b} and \eqref{230405c}, we see that $\At_a / \At_g \propto a^3$.

Gravitational tides tend to circularise the orbits, but usually on a much longer timescale than they take to modify the spin \citep[e.g.][]{Valente_Correia_2022}.
Moreover, if the system has several planets, their mutual gravitational interactions permanently excite the eccentricities \citep[e.g.][]{Laskar_1994, Laskar_2008}.
Therefore, we expect that Earth-like planets can be observed with some eccentricity, provided that the system remains stable.
For eccentric orbits (Fig.~\ref{Fig:4.1}, orange and blue curves), the number of possible equilibria for the spin increases and the figure becomes more complex.
In this case, gravitational tides alone can already drive the rotation rate into asynchronous spin-orbit resonances \citep[e.g.][]{Makarov_Efroimsky_2013, Correia_etal_2014}.
These new states centred at $ \om/n = k/2 $ (with $k \in \mathbb{Z}$) arise from the contribution of the many Hansen coefficients (Eq.\,\eqref{eqS447} and Tab.~\ref{tabHansen}).
Indeed, higher-order spin-orbit resonances appear as we increase the eccentricity.
When we take into account thermal tides, additional nonzero Hansen coefficients are present at the exact same frequencies (Eq.\,\eqref{eqS447}), but do not change much the balance of torques.
Gravitational tides thus control the spin evolution near the spin-orbit resonances, which remain a stable possible outcome for the spin evolution.
Nevertheless, as in the case with zero eccentricity, thermal tides can still introduce an additional asynchronous and non resonant state at both ends of the spin-orbit resonance region, although these equilibria become more difficult to attain (the critical semi-major axis $a_c$ increases with the eccentricity).
In the example shown in Fig.~\ref{Fig:4.1}\,c (with $a=0.35$~au), for $e=0.2$, the thermal asynchronous equilibria are still present, but for $e=0.4$, they are no longer possible.

\subsection{Asynchronous non-resonant thermal states}
\label{ss:anrts}

During the formation stages, planets are expected to acquire fast rotation rates \citep[e.g.][]{Kokubo_Ida_2007}, such that we initially have $|\om| \gg n$.
Then, for zero obliquity, as the rotation approaches the equilibria region ($\om \sim n$), it first encounters and locks in one of the asynchronous non-resonant equilibria introduced by thermal tides (provided that these states exists for a given semi-major axis and eccentricity).
Since asynchronous thermal states are of major interest for habitability, it is important to better understand how they depend on the orbital parameters.

\begin{figure}
\centering
\includegraphics[width=\columnwidth]{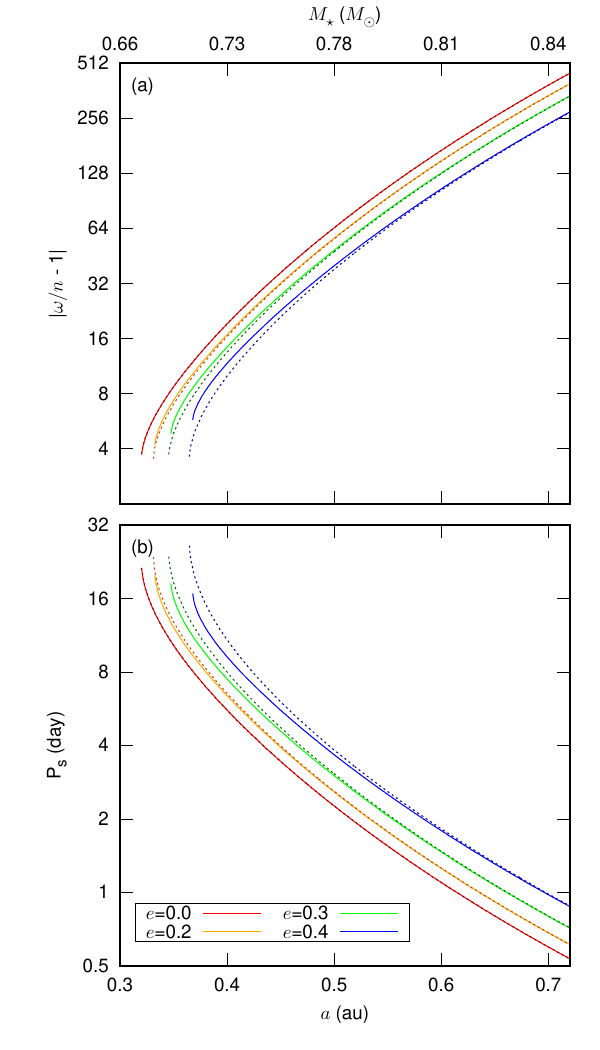}
\caption{Relative synodic frequency, $\om_s/n = |\om/n - 1|$ (top), and respective synodic rotation period, $P_s = 2 \pi / \om_s$ (bottom), of the stable asynchronous thermal equilibrium states as a function of the semi-major axis for different values of the eccentricity, $e=0, 0.2, 0.3, 0.4$. The solid lines give the state with $\om/n > 1$ (super-synchronous), while the dashed lines give the state with $\om/n < 1$ (sub-synchronous). These plots are shown on a log-scale in the y-axis.}\label{Fig:4.2} \end{figure}

In Fig.~\ref{Fig:4.2}, we first determine the position of the asynchronous thermal states as a function of the semi-major axis, for some fixed values of the eccentricity ($e=0$, $e=0.2$, $e=0.3$, and $e=0.4$).
The different equilibria are obtained numerically by solving $\dot \om = 0$ (Eq.\,\eqref{eqS447}).
We show the corresponding relative synodic frequency, %\footnote{\bf In the case of the Earth, the Lamb resonance occurs at $\om_s/n \approx 263$ \citep{Farhat_etal_2023}, which may impact the results in Fig.~\ref{Fig:4.2} for $a > 0.7$.}
$\om_s/n = |\om/n - 1|$, but also the synodic rotation period, $P_s = 2 \pi / \om_s$ (in days).
The solid lines follow the state with $\om/n > 1$ (super-synchronous), while the dashed lines map the state with $\om/n < 1$ (sub-synchronous).
The synodic frequency is the same for zero eccentricity \citep{Correia_Laskar_2001}, but it can be slightly different for eccentric orbits, in particular close to the critical semi-major axis.
We recall that in our analysis we fix the stellar flux at the same value as that of the Earth's.
Therefore, by modifying the semi-major axis, we also change the stellar mass according to expressions \eqref{eqS446} and \eqref{eqS444}.
The mass range $0.20 < \ms /  M_\odot < 0.85 $ corresponds to a semi-major axis range $0.07 < a < 0.72$~au.

In Fig.~\ref{Fig:4.2}, we observe that the relative equilibrium rotation rate, $\om/n$, increases with the semi-major axis (and hence with the stellar mass).
Conversely, the synodic rotation period decreases with the semi-major axis, that is, the planet rotates faster.
Interestingly, we find that the equilibrium can be close to $P_s \approx 1$~day (the current synodic period of the Earth), for some semi-major axes.
More precisely, for $a \approx 0.62$~au when $e=0$ (circular orbit), for $a \approx 0.64$~au when $e=0.2$ (small eccentricity), for $a \approx 0.66$~au when $e=0.3$, and for $a \approx 0.70$~au when $e=0.4$ (moderate eccentricity).
These semi-major axes provide the same stellar flux of the Earth for a star with $\ms \approx 0.82 \, M_\odot$. 
We hence conclude that Earth-like planets around low-mass stars can maintain a rotation period similar to that of the current Earth despite being closer to their parent stars.

In Fig.~\ref{Fig:4.2}, we can also clearly see that there is a critical semi-major axis below which the asynchronous thermal states no longer exist. 
Planets with semi-major axes below the critical values thus evolve into some spin-orbit resonance (Fig.~\ref{Fig:4.1}).
We further confirm that the critical semi-major axes depend on the eccentricity.
As the eccentricity increases, the critical value of the semi-major axis also increases (check also Fig.~\ref{Fig:4.1}\,c).
Moreover, for eccentric orbits, we can spot a small difference in the critical semi-major axes between the sub- and super-synchronous equilibrium states.
The equilibrium rotations between these two states also differ more significantly close to the critical semi-major axes.
%In particular, we observe that the sub-synchronous state may be still present, while the  super-synchronous state is no longer possible.
These subtle differences become more visible for the case with higher eccentricity ($e=0.4$).

\begin{figure}
\centering
\includegraphics[width=\columnwidth]{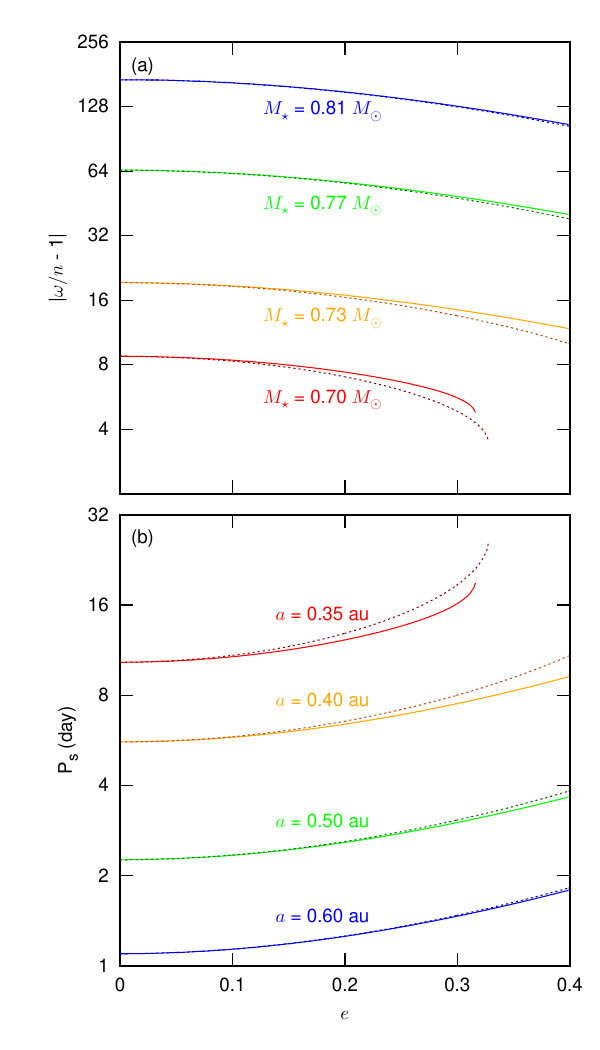}
\caption{Relative synodic frequency, $\om_s/n = |\om/n - 1|$ (top), and respective synodic rotation period, $P_s = 2 \pi / \om_s$ (bottom), of the stable asynchronous thermal equilibrium states as a function of the eccentricity for different values of the semi-major axis, $a= 0.35, 0.4, 0.5, 0.6 $~au, corresponding to the stellar masses $\ms / M_\odot \approx  0.70, 0.73, 0.77, 0.81 $, respectively. The solid lines give the state with $\om/n > 1$ (super-synchronous), while the dashed lines give the state with $\om/n < 1$ (sub-synchronous). These plots are shown on a log-scale in the y-axis.}\label{Fig:4.3}
\end{figure}

Since the eccentricity introduces some differences in the final equilibria, in Fig.~\ref{Fig:4.3}, we determine the position of the asynchronous thermal states as a function of the eccentricity, for some fixed values of the semi-major axis from $a=0.35 $~au to $0.6$~au.
All these semi-major axis values are larger than the critical semi-major axis for a circular orbit.
Because we fix the stellar flux, the stellar masses are more or less\footnote{Actually, the stellar flux also changes with the eccentricity (Eq.\,\eqref{eqS446}), and hence the stellar mass. However, the impact of the eccentricity is not very important, because for the maximum value chosen ($e=0.4$), we still have $\sqrt{1-e^2} \approx 0.92$.} constant for each semi-major axis, going from $\ms / M_\odot \approx 0.7$ to $0.8$, respectively.
We show again the relative synodic frequency, $\om_s/n = |\om/n - 1|$, together with the corresponding synodic rotation period, $P_s = 2 \pi / \om_s$ (in days).
The solid lines show the state with $\om/n > 1$ (super-synchronous), while the dashed lines give the state with $\om/n < 1$ (sub-synchronous).
We confirm that the synodic frequency is the same in the two states for zero eccentricity, but the difference increases with the eccentricity, in particular close to the critical semi-major axis.
We also observe that the effect of the eccentricity is more important for smaller semi-major axis.
For $a=0.35$~au, the critical semi-major axis is attained for $e\approx0.33$, which explains why no asynchronous thermal equilibria is observed for $e=0.4$ in Fig.~\ref{Fig:4.1}\,c.

\subsection{Habitable zone} 
\label{HZ}

In the previous section, we have seen that for Earth-like planets, asynchronous thermal states can only exist for semi-major axes larger than some critical value, $ a_c \approx 0.35$~au, which correspond to stars with $\ms / M_\odot \gtrsim 0.7 $ (K-dwarf stars).
However, this analysis was restricted to an exact Earth-twin planet, that is, a planet with the same mass, radius, atmosphere composition, and average annual stellar flux as the Earth's (Tab.~\ref{tab:venus}).

The existence of life as we know it, based on liquid water, is likely not so restrictive. 
In principle, habitable worlds can exist for a board range of stellar fluxes, $450 < F_\osdot < 1508$~Wm$^{-2}$, known as habitable zone \citep[e.g.][]{Kasting_etal_1993, Leconte_etal_2015}.
The amplitude of thermal tides depends on the stellar flux (Eq.\,\eqref{eqS232}), so the position of the asynchronous thermal states will also change.
For an Earth-like planet (N$_2$ atmosphere and $p_s = 1$~bar), \citet{Leconte_etal_2015} also compute the thermal response for the stellar flux $F_\osdot = 450$~Wm$^{-2}$ using general circulation models.
They obtain for the radiative frequency of the atmosphere, $\sigma_0 = 74.5$~yr$^{-1}$, % $\sigma_0 = 2 \om_0$
and for the amplitude of the atmospheric quadrupole, $q_0 = 8.90$~mbar.
It is then possible to determine the position of the different equilibria for the spin for this specific stellar flux.
We can also determine the critical semi-major axis. 
For a circular orbit we get $a_c = 0.34$~au, which corresponds to a stellar mass $\ms / M_\odot = 0.51$.
This critical semi-major axis is slightly larger than the one obtained for Earth's stellar flux (Fig.~\ref{Fig:4.1}\,b), but the stellar mass is much smaller and already in the M-dwarf regime.
Since the critical semi-major axes appears to not be very sensitive to the stellar flux, from the two points provided by \citet{Leconte_etal_2015} at $F_\osdot = 450$~Wm$^{-2}$ and $1366$~Wm$^{-2}$, one can extrapolate using a straight line that, for Earth-like planets, we have
\be
%a_c = - 2.29 \times 10^{-5} F_\osdot + 0.351 \ \mathrm{au} 
%a_c \approx - 2.3 \times 10^{-5} F_\osdot + 0.35 \ \mathrm{au} 
%
%a_c = (0.4 - 0.117647 \, \ms) \, \mathrm{au} 
a_c \approx (0.4 - 0.12 \, \ms / M_\odot) \, \mathrm{au} 
\label{230516a} \ .
\ee

\begin{figure}
\centering
\includegraphics[width=\columnwidth]{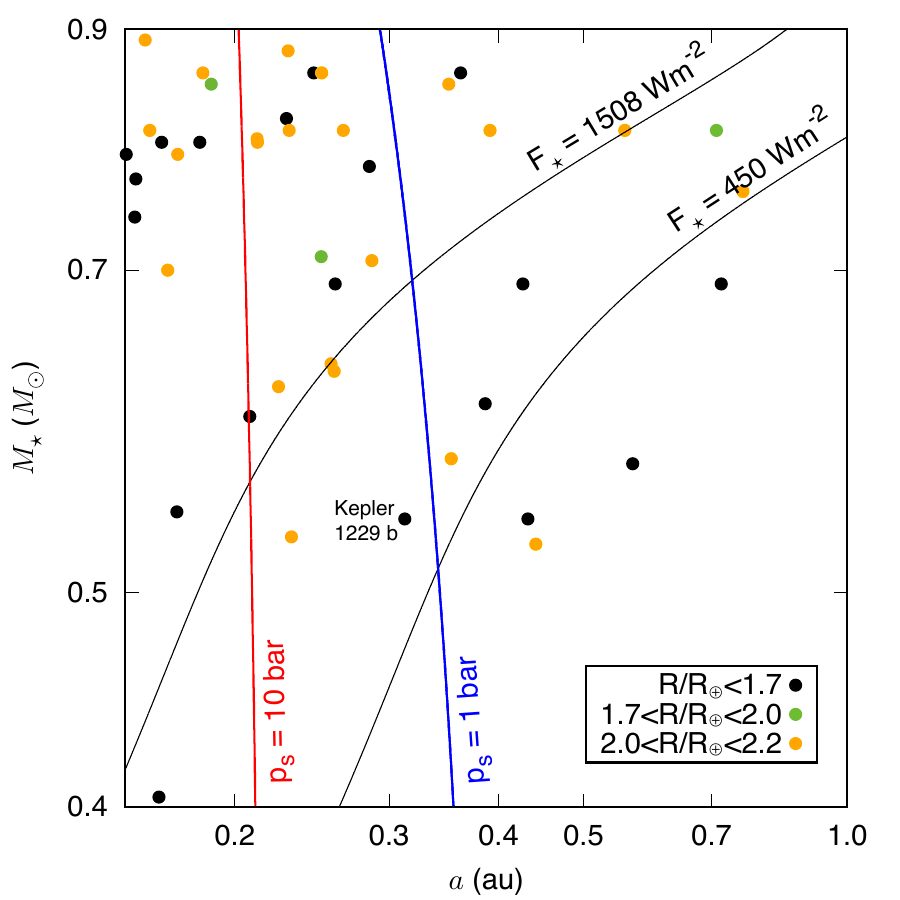}
\caption{Critical semi-major axes for Earth-like planets around low-mass stars. We show the results for an atmosphere with $p_s = 1$~bar (in blue) and $p_s = 10$~bar (in red). Planets on the right-hand side of the critical lines are expected to present asynchronous rotation. We also show the limits of the habitable zone (black curves) for stellar fluxes within $450 < F_\osdot < 1508$~Wm$^{-2}$. The dots correspond to the already known planets (in black for $\Rp / R_{\oplus} <1.7 $, in green for $1.7 < \Rp / R_{\oplus} < 2.0 $, and in orange for $2.0 < \Rp / R_{\oplus} < 2.2$).  } \label{Fig:4.4}
\end{figure}

In order to get a more comprehensive view of the possible behaviour of Earth-like planets at different stellar fluxes, in Fig.~\ref{Fig:4.4}, we show the critical semi-major axes (Eq.\,\eqref{230516a}) as a function of the stellar mass and semi-major axis for circular orbits (blue line).
Only planets on the right-hand side of this line can be found in an asynchronous thermal state, while those on the left-hand side will likely develop synchronous rotations (at least for zero eccentricity).
In Fig.~\ref{Fig:4.4}, we also show the limits of the habitable zone (black lines), to better guide us to the most interesting areas.
The top line corresponds to $F_\osdot = 1508$~Wm$^{-2}$, while the bottom line corresponds to $F_\osdot = 450$~Wm$^{-2}$.
We observe that for $\ms / M_\odot \gtrsim 0.5 $ and $ a \gtrsim 0.35$~au, it is possible to find a wide region where planets can be simultaneously asynchronous and inside the habitable zone\footnote{Our Fig.~\ref{Fig:4.4} is similar to Fig.~3 in \citet{Leconte_etal_2015}, but it has some important differences for the critical semi-major axes.
They used the constant$-Q$ model for gravitational tides, while we adopted the Andrade model (Sect.~\ref{sect:gtt}).
More importantly, in Eq.\,\eqref{230516a} we compute the mass of the star that is able to provide a given stellar flux $F_\osdot (a_c, \ms) $ according to expressions \eqref{eqS446} and \eqref{eqS444}.
%For the equivalent Eq.\,(2) in \citet{Leconte_etal_2015}, it appears that they adopt a fixed stellar flux and then $a_c$ is just proportional to the stellar mass.
}.

Habitable worlds can also exist for a large variety of planetary properties \citep[e.g.][]{Kopparapu_etal_2013, Kopparapu_2019}, that is, they are not restricted to the Earth's mass, radius, and atmosphere.
Actually, because of the current detection limitations, most of the low-mass (rocky) planets known to date are in the super-Earth regime \citep[e.g.][]{Schlichting_2018}.
For comparison, in Fig.~\ref{Fig:4.4}, we plot all the known planets with $\Rp / R_\oplus < 2.2 $ (which is equivalent to $\mp / M_\oplus < 10$, assuming a mean density identical to that of the Earth's).
For planets with $\Rp / R_\oplus < 1.7 $ (that is, with $\mp / M_\oplus < 5$, marked with a black dot), the mass is not large enough to accrete a voluminous gaseous envelope, so they are likely composed of a rocky mantle and a thin atmospheric layer, as observed for Earth and Venus \citep[e.g.][]{Demory_etal_2016, Komacek_Abbot_2019, Wordsworth_Kreidberg_2022}.

For rocky planets in the super-Earth regime, we nevertheless expect a more dense atmosphere than that of the Earth.
Using general circulation models, \citet{Auclair-Desrotour_etal_2019a} derived some scaling laws for $q_0$ and $\sigma_0$ as a function of the semi-major axis, for a dry terrestrial planet with a homogeneous N$_2$ atmosphere and $p_s = 10$~bar around a Sun-like star\footnote{From Table~2 in \citet{Auclair-Desrotour_etal_2019a}, the scaling laws are $\log_{10} q_m  = -0.69 \log_{10} a + 3.25$, and $\log_{10} \tau_m = 0.86 \log_{10} a + 0.48$, where $a$ is given in au, $q_m$ in Pa, and $\tau_m$ in day, $q_0 = 2 q_m$ and $\sigma_0 = 1/ \tau_m$.}.
These laws can be easily converted in terms of stellar flux as (Eq.\,\eqref{eqS446})
\be 
\log_{10} a  = 1.568 - \frac12 \log_{10} F_\osdot \ ,
\ee
which gives
%%%%%%%%%%
%\be \log_{10} q_m  = 0.345 \log_{10} F_\osdot + 2.168 \ee
%\be \log_{10} \tau_m =  - 0.430 \log_{10} F_\osdot + 1.828 \ee
%%%%%%%%%%
\be
\log_{10} q_0 = 0.345 \log_{10} F_\osdot + 2.469 
\label{230607q}
\ee
and
\be
\log_{10} \sigma_0  =  0.430 \log_{10} F_\osdot + 0.7343 \ ,
\label{230607s}
\ee
where $a$ is in au, $F_\osdot $ is given in Wm$^{-2}$, $q_0$ in Pa, and $\sigma_0$ in yr$^{-1}$.
Therefore, it is possible to know the thermal tidal response of the atmosphere for any planet of this kind, and hence the position of the asynchronous thermal states.

In Fig.~\ref{Fig:4.4}, we additionally show the critical semi-major axes for an Earth-like planet with $p_s = 10$~bar (red line).
Interestingly, we observe that the critical semi-major axis and the stellar mass required to prevent synchronous rotation are much smaller than for a planet with $p_s = 1$~bar (blue line).
Many of the already known planets are in the right-hand side of this critical line, so we expect them to present asynchronous rotation.
Some of these planets are also inside the habitable zone, which can foster friendly conditions for life to develop.

%%%%%%%%%%

\section{Application to Kepler-1229\,b}
\label{apliKepler}

%As for Venus in Sect.~\ref{ssct:cse}, 
In order to study the complete long-term tidal evolution of a given Earth-like system over time, we need to attribute specific values to all physical and dynamical parameters.
One good candidate for habitability studies is Kepler-1229\,b \citep{Morton_etal_2016}, a super-Earth with a radius of $\Rp = 1.40 \, R_\oplus$ in a $87$~day orbit around a star with mass $\ms = 0.54 \, M_\odot$ ($a = 0.3125$~au), which is in the middle of the habitable zone  (Fig.~\ref{Fig:4.4}).
The eccentricity is not constrained because so far this system was solely observed using the transits technique.

Despite Kepler-1229\,b being slightly larger than Earth, we adopted here a similar average density, inner structure constant, Love numbers, relaxation time, and Andrade rheology.
As a result, we obtained a larger mass, $\mp \approx 2.74 \, M_\oplus$, which is able to accrete more gas. 
%{The stellar flux received by this planet is also lower compared to the Earth, requiring a thicker atmosphere, too} \citep{Ohja_etal_2022}.   
Therefore, we assumed a denser N$_2$ atmosphere for Kepler-1229\,b, with $p_s = 10$~bar. 
For a circular orbit,\footnote{For $e\ne0$, we still use Eq.\,\eqref{eqS446}, but we get a slightly different mean flux, and thus also slightly different values for $q_0$ and $\sigma_0$.} we get from Eq.\,\eqref{eqS446} that $F_\osdot \approx 606$~Wm$^{-2}$, which in turn allowed us to obtain for the amplitude of the atmospheric quadrupole, $q_0 = 26.9$~mbar (Eq.\,\eqref{230607q}), and for the radiative frequency of the atmosphere, $\sigma_0 = 85.3$~yr$^{-1}$ (Eq.\,\eqref{230607s}). % $\sigma_0 = 2 \om_0$
In Table~\ref{tab:venus}, we summarise the physical and dynamical parameters that we used.

\subsection{Equilibrium final states}

\begin{figure}
\centering
\includegraphics[width=\columnwidth]{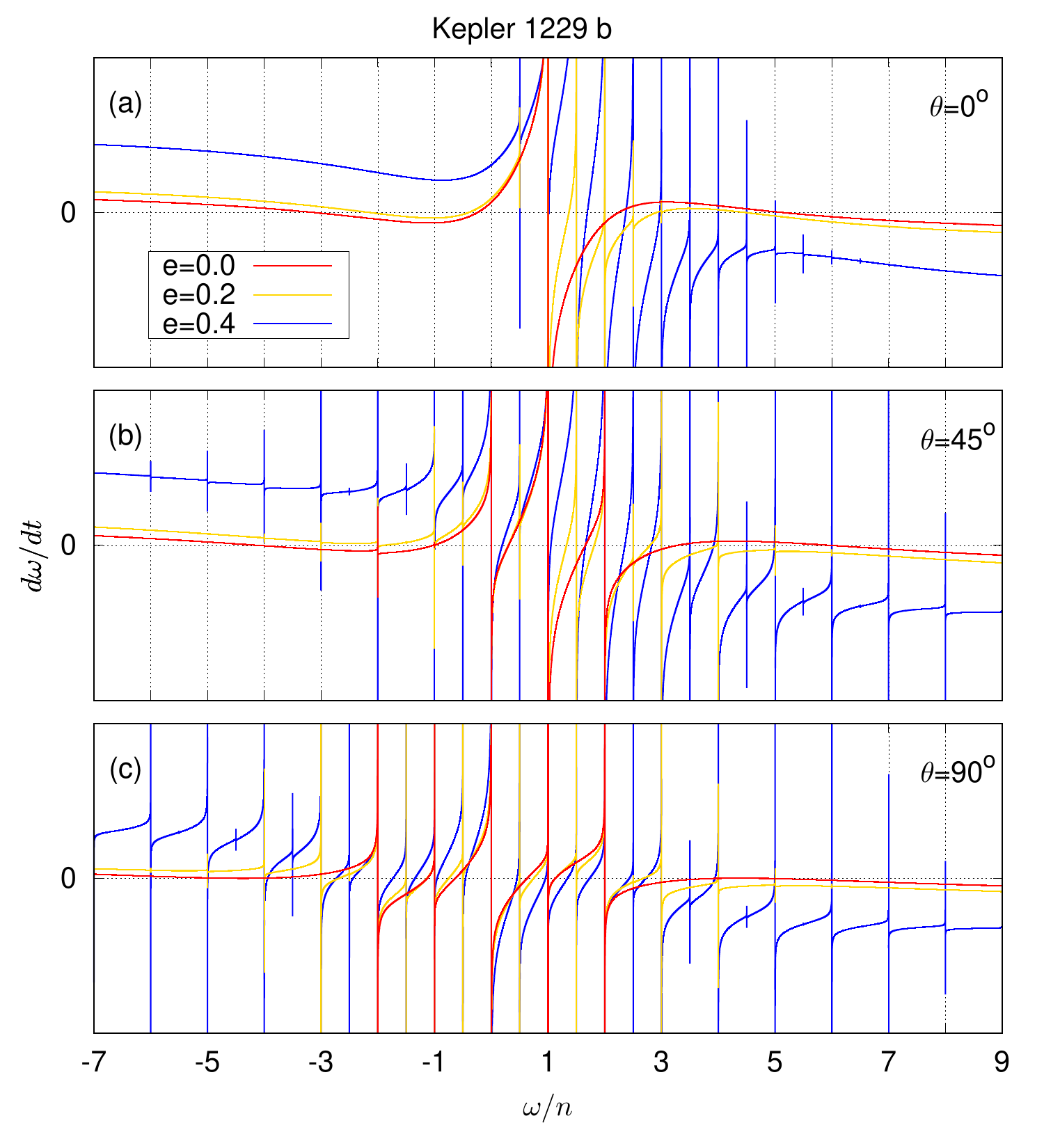}
\caption{Variation of $d \om/ dt $ upon $\om / n $ for Kepler-1229\,b. We show the evolution for different obliquity values: $\theta = 0^\circ$ (top), $\theta = 45^\circ$ (middle), and $\theta = 90^\circ$ (bottom plot). We also show the evolution for three different values of the eccentricity: $e=0.0$ (red line), $e=0.2$ (orange line),  and $e=0.4$ (blue line).} \label{Fig:5.1}
\end{figure}

We first investigate the possible equilibrium final states for the given parameters of Kepler-1229\,b.
In Fig.~\ref{Fig:5.1}\,a, we show the evolution of the rotation rate for zero obliquity (Eq.\,\eqref{eqS447}) and different values of the eccentricity ($e = 0$, $e=0.2$, and $e=0.4$).
As for the case of Venus (Sect.~\ref{ssct:es}) and for an Earth-like planet with $a>a_c$ (Sect.~\ref{efs:earth}), we observe that for zero eccentricity there are three stable equilibrium points, one for $\om = n$ (synchronous rotation), and another two for $\om = n \pm \om_s$ (asynchronous thermal states).
More precisely, we find $\om_s/n \approx 4.1$ (corresponding to $P_s \approx 21 $~day), which gives one prograde state at $\om/n \approx 5.1$, and one retrograde state at $\om/n \approx -3.1$.
%We conclude that for Kepler-1229\,b, we also have $a>a_c$ (see also Fig.~\ref{Fig:4.4}).
As we increase the eccentricity, additional equilibria appear at spin-orbit resonances $\om/n = k/2 $ (with $k \in \mathbb{Z}$), while the asynchronous thermal states tend to disappear.
They still exist for $e = 0.2$ (for $\om/n \approx 4.5$ and $\om/n \approx -2.2$), but they are no longer possible for $e=0.4$.
%In this last case, the most likely spin-orbit resonance to be captured in is for $\om / n = 4$ (for planets evolving from fast rotation rates).

In Fig.~\ref{Fig:5.1}\,b, we show the evolution of the rotation rate for an obliquity $\theta = 45^\circ$, while in Fig.~\ref{Fig:5.1}\,c, we show the evolution for  $\theta = 90^\circ$.
We observe that, as we increase the obliquity, the presence of asynchronous thermal states becomes more difficult.
They are still present for $\theta = 45^\circ$ in the circular case ($e=0$), but for $\theta = 90^\circ$ they are no longer possible for any eccentricity value.
On the other hand, for nonzero obliquities, the rotation rate can be trapped in additional spin-orbit resonances, even in the circular case, for $\om/n = -2,-1,0,+1,+2$, as it was already reported for Venus (Fig.~\ref{Fig:3.2}).
We hence conclude that capture in the asynchronous thermal states is more likely for low obliquity and small eccentricity, while large obliquities and/or high eccentricities favour the capture in spin-orbit resonances.

\subsection{Numerical simulations}

We now simulate the complete tidal evolution of Kepler-1229\,b, by integrating the full set of secular equations of motion for 10~Gyr (Eqs.\,\eqref{150626a}, \eqref{210805b}, and \eqref{211110b}). 
The initial spin state of Earth-like planets is unknown, since large impacts during the formation process can completely modify their rotation period and spin axis orientation \citep{Dones_Tremaine_1993, Kokubo_Ida_2007}.
In all simulations we assumed that the initial rotation period of Kepler-1229\,b is one day, which gives an initial rotation rate of $\om / n \approx 87$.
Since the rotation evolves into an equilibrium state, the choice of its initial value is not critical.
For the initial obliquity, we adopted different values between $0^{\circ}$ and $180^{\circ}$.   
 
Despite the fact that we take into account the orbital evolution in our model (Eqs.\,\eqref{150626a} and \eqref{211110b}), we observed that the semi-major axis and the eccentricity remain approximately constant throughout the evolution, which is due to $||\vG|| \gg ||\vL||$, where $\vG$ and $\vL$ are the orbital and rotational angular momentum, respectively (Eqs.\,(\ref{eqS21}) and (\ref{eqS22})).
Therefore, we assume for the initial semi-major axis the present value, $a = 0.3125$~au. 
The eccentricity of Kepler-1229\,b is unknown, but this parameter is critical for the spin evolution because it can lead to different final states (Fig.~\ref{Fig:5.1}).
To explore the multiple evolution possibilities, we then adopt three different values for the initial eccentricity, $e=0$, $e=0.2$ and $e=0.4$.

\begin{figure}
\centering
\includegraphics[width=\columnwidth]{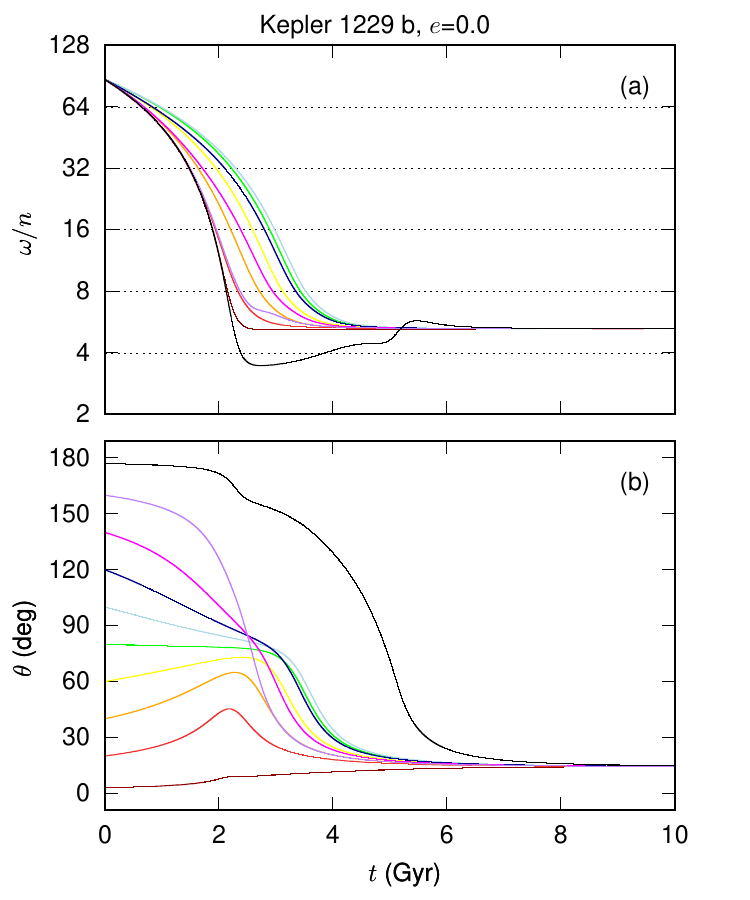}
\caption{Secular evolution over time for Kepler $1229$ b with $e=0.0$ and for initial obliquity values between $0^\circ$ and $180^\circ$. We show the $\om /n $ ratio (top) and the obliquity (bottom). In all simulations, the initial rotation period is 24~hours ($\om / n \approx 87$).} \label{Fig:5.2}
\end{figure}

\begin{figure}
\centering
\includegraphics[width=\columnwidth]{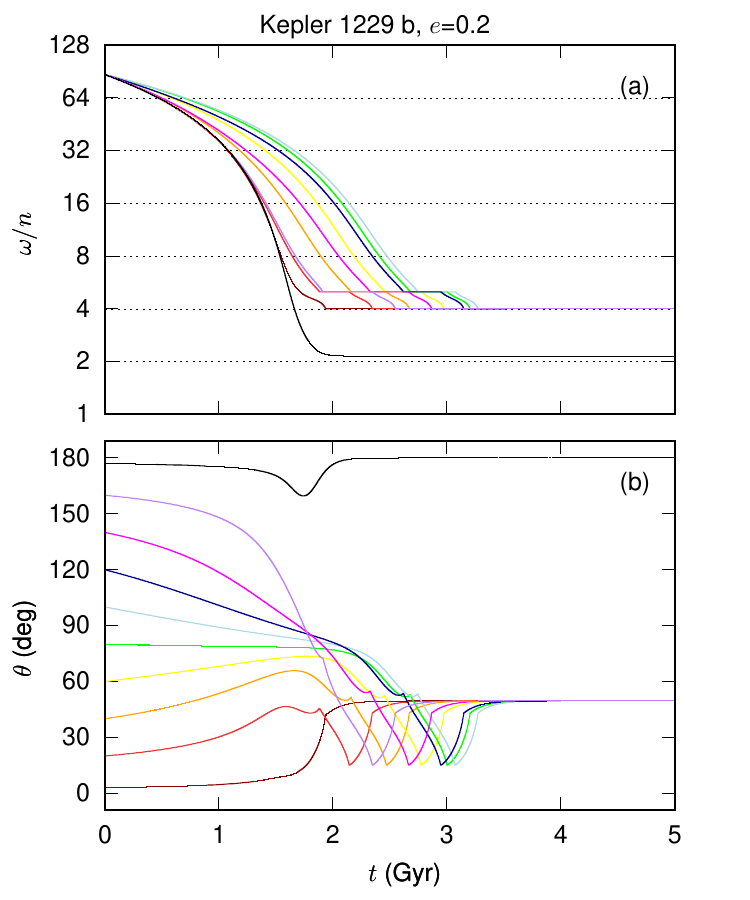}
\caption{Secular evolution over time for Kepler $1229$ b with $e=0.2$ and for initial obliquity values between $0^\circ$ and $180^\circ$. We show the $\om /n $ ratio (top) and the obliquity (bottom). In all simulations, the initial rotation period is 24~hours ($\om / n \approx 87$).} \label{Fig:5.3}
\end{figure}

\begin{figure}
\centering
\includegraphics[width=\columnwidth]{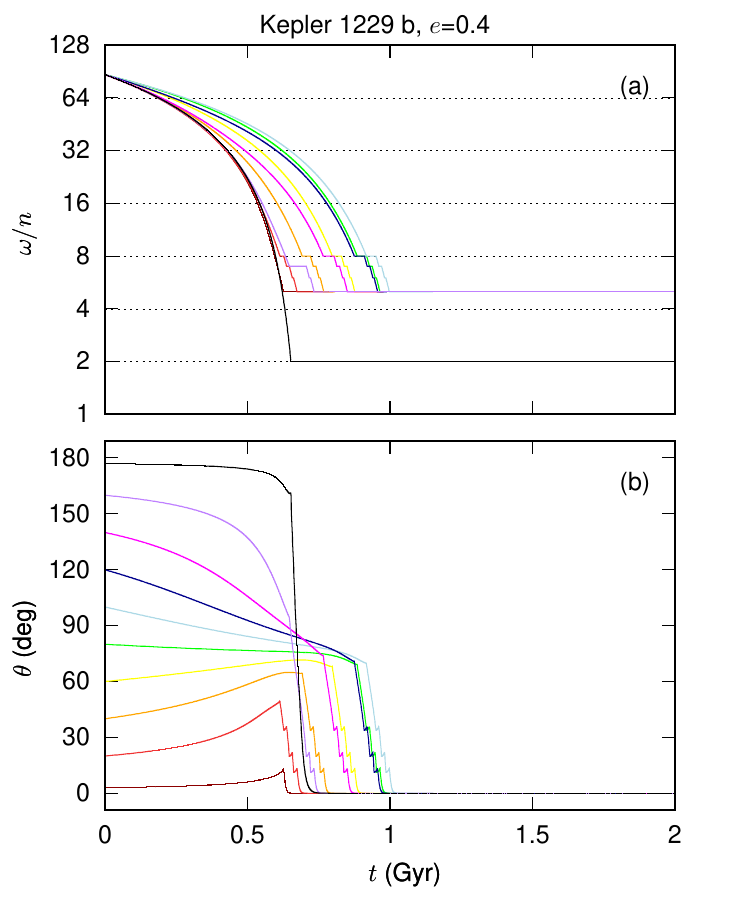}
\caption{Secular evolution over time for Kepler $1229$ b with $e=0.4$ and for initial obliquity values between $0^\circ$ and $180^\circ$. We show the $\om /n $ ratio (top) and the obliquity (bottom). In all simulations, the initial rotation period is 24~hours ($\om / n \approx 87$). } \label{Fig:5.4}
\end{figure}

The results of the numerical simulations for the tidal evolution of Kepler-1229\,b over time are shown in Figs.~\ref{Fig:5.2} to~\ref{Fig:5.4}, one for each initial eccentricity value.
Since the orbital evolution is negligible for this system, we focus our analysis in the spin evolution of the planet (Eq.\,\eqref{210805b}).
We thus only show the $\om/n$ ratio (top) and the obliquity (bottom) as a function of time for several different initial obliquities.
%Because we initially have $\om/n \gg 1$, the rotation rate evolution while approaching the equilibrium states  ($\om / n \sim 1$) can be better understood by comparing with Fig.~\ref{Fig:5.1} for decreasing values of $\om/n$.

In Fig \ref{Fig:5.2}, we show the evolution for $e=0$ (circular orbit). 
%We observe that, regardless of the initial obliquity, all trajectories end in an asynchronous thermal state in less than about 5~Gyr. 
%The equilibrium final rotations are in good agreement with the equilibrium points predicted in Fig.~\ref{Fig:5.1}\,a (red line).
%For initial obliquities very close to $180^\circ$, the spin evolves into the retrograde state $\om/n = \om_s/n - 1 \approx 3.1$, with $\theta = 180^\circ$. % w/n = 3.13
%For the remaining initial obliquities, the spin ends in the prograde state $\om/n = \om_s/n + 1 \approx 5.2$ with low obliquity, but not exactly zero ($\theta \approx 14^\circ$). % w/n = 5.23
%For initial obliquities very close to $0^\circ$, the obliquity does not have time to attain the nonzero equilibrium value in less than 10~Gyr, but it also converges into $\theta \approx 14^\circ$ if we extend the simulation length.
We observe that, regardless of the initial obliquity, all trajectories end in the prograde asynchronous thermal state in less than about 5~Gyr. 
More precisely, the spin ends in the prograde state $\om/n = \om_s/n + 1 \approx 5.2$ with low obliquity, but not exactly zero ($\theta \approx 14^\circ$). % w/n = 5.23
The nonzero final obliquity is somehow surprising, because in the circular case, gravitational and thermal tides usually push the obliquity into either $0^\circ$ or $180^\circ$, respectively \citep{Correia_etal_2003}.
Of course, in this case the two effects are simultaneously in action, so both torques can cancel for some obliquity (as it happens for the rotation rate).    

In Fig. \ref{Fig:5.3}, we show the evolution for $e=0.2$ (small eccentricity).
We observe that all simulations end in a stable equilibrium in less than about 4~Gyr.
However, contrarily to the circular case, depending on the initial obliquity, in this case the rotation can be either captured in an asynchronous thermal state, or in some spin-orbit resonance.
%For initial obliquities very close to $0^\circ$ and $180^\circ$, the spin evolves into the asynchronous thermal states ($\om/n \approx 4.4$, $\theta = 0^\circ$) and ($\om/n \approx 2.1$, $\theta = 180^\circ$), respectively.
%On the other hand, for initial obliquities in between ($\theta \in [20^\circ, 160^\circ]$), the rotation rate stabilises in the $\om/n = 4/1$ spin-orbit resonance, while the obliquity settles at $\theta \approx 50^\circ$.
Indeed, in Fig.~\ref{Fig:5.1}\,b, we observe that for $e=0.2$ and moderate obliquity, the asynchronous thermal states are no longer possible.
For initial obliquities close to $180^\circ$, the spin evolves into the retrograde asynchronous thermal state $\om/n \approx 2.2$ with $\theta = 180^\circ$. % w/n = 2.15
On the other hand, for initial obliquities lower than $160^\circ$, the rotation rate stabilises in the $\om/n = 4/1$ spin-orbit resonance, while the obliquity settles at $\theta \approx 50^\circ$. % ob = 49.6
Indeed, in Fig.~\ref{Fig:5.1}\,b, we observe that for $e=0.2$ and an obliquity of $45^\circ$, the asynchronous thermal states are no longer possible.
We also observe that the first spin-orbit resonance where capture can occur is for $\om/n=5/1$.
That is why in our simulations the rotation is initially captured in this resonance (Fig.~\ref{Fig:5.3}).
However, in the 5/1 resonance the obliquity decreases. 
When the obliquity descends below the threshold $\theta \approx 17^\circ$, the resonant equilibrium is no longer possible, and the rotation rate decreases.
Then, the obliquity increases again and the rotation evolves into the following resonance, $\om/n=4/1$, where it can be trapped permanently because the obliquity remains high (Fig.~\ref{Fig:5.1}\,b).
As in the circular case, it is remarkable that such high obliquities can be maintained. 
\citet{Valente_Correia_2022} have shown that, when considering gravitational tides alone, some half-integer spin-orbit resonances can excite the obliquity to high values.
Here we find that when we additionally include thermal tides, integer resonances can also have the same kind of effect.

In Fig. \ref{Fig:5.4}, we show the evolution for $e=0.4$ (moderate eccentricity).
We observe that all simulations end in a stable equilibrium in less than about 1~Gyr.
In this case, asynchronous thermal equilibria are not possible for any obliquity value (Fig.~\ref{Fig:5.1}), so the rotation must evolve into a spin-orbit resonance.
Indeed, regardless of the initial obliquity, we observe a succession of resonance captures over time in Fig.~\ref{Fig:5.4}. 
For initial obliquities close to $180^\circ$, the rotation is directly captured in the $\om/n=2/1$ spin-orbit resonance, and the final obliquity evolves into $\theta =0^\circ$.
On the other hand, for initial obliquities lower than $160^\circ$,
the obliquity first evolves into a high value around $\theta=75^\circ$, which facilitates the capture in high order spin-orbit resonances, such as $\om/n = 8/1$.
However, as soon as capture occurs, the obliquity decreases and the spin-orbit resonances are exited after the obliquity descends below a certain limit.
The last stage for the spin evolution occurs when all trajectories are captured in the resonance $\om/n=5/1$, where the final obliquity also evolves into $\theta =0^\circ$.
This spin-orbit resonance was previously unstable at low obliquity for $e=0.2$, but it is the first resonance to become stable when $e=0.4$ (Fig.~\ref{Fig:5.1}\,a).
We conclude that the spin evolution for $e=0.4$ is quite similar to the spin evolution in absence of thermal tides \citep[see][]{Valente_Correia_2022}.
This result is not totally unexpected, because for higher eccentricities the critical semi-major axis is larger (Fig.~\ref{Fig:4.2}), and for $a<a_c$, thermal tides are dominated by gravitational tides (Fig.~\ref{Fig:4.4}).
%Since the eccentricity enhances gravitational tides at the pericentre, gravitational tides become dominating rather for $a (1-e) <a_c$.
%As we increase the eccentricity, the evolution timescale decreases, because gravitational tides become stronger at the pericentre.

\section{Discussion and conclusions}
\label{conclusion}

In this paper, we have revisited the long-term evolution of Earth-like planets while taking into account both gravitational and thermal atmospheric tides.
Preliminary studies have shown that the combined effect of these two kinds of tides may lead to asynchronous equilibrium states, which can help sustain temperate environments and thus more favourable conditions for life \citep{Correia_etal_2008, Cunha_etal_2015, Leconte_etal_2015, Revol_etal_2023}.
Here, we adopted a vectorial formalism based on Hansen coefficients expansions \citep{Correia_Valente_2022, Valente_Correia_2023},
which allowed us to easily follow the spin and the orbital evolution of these planets for any obliquity and orbital eccentricity.
This formalism is very general and can be used with a large variety of rheologies and atmospheric models.

Venus is believed to be a unique example in the Solar System of a planet with a balanced equilibrium between gravitational and thermal tides \citep[e.g.][]{Gold_Soter_1969, Dobrovolskis_1980, Correia_Laskar_2001, Auclair-Desrotour_etal_2017a}.
If this is so, it can therefore be used to gauge some of the unknown parameters in our models.
When adopting an Andrade rheology for gravitational tides \citep{Andrade_1910}, we found that the time exponent $\alpha = 0.3$ and a Maxwell relaxation time of $\tau \approx 1500$~yr provide the best fit for the observations.
Although these parameters were obtained from dynamical constraints on the present rotation of Venus, they are consistent with those commonly derived for the Earth and rocky planets using alternative methods \citep[e.g.][]{Castillo_Rogez_etal_2011, Tobie_etal_2019, Bolmont_etal_2020}.

When we applied our model to Earth-like planets, we confirmed that asynchronous rotation can indeed be expected for planets in the habitable zone of low-mass stars ($0.4 < \ms / M_\odot < 0.85$).
Interestingly, we found that Earth-like planets in the habitable zone of stars with masses $\sim 0.8$~$M_\odot$ may end up with an equilibrium rotation period of 24~hours.
However, these planets require semi-major axes around 0.6~au, so the evolution into this equilibrium state is slow; it can take the entire age of the system to achieve depending on the initial rotation rate of the planet.
Conversely, if the initial rotation period is already close to 24~hours, it will not evolve much.
This result is very compelling but needs to be taken with caution because the equilibrium rotations are very sensitive to the models and parameters adopted for the atmosphere. 
For instance, we did not consider the effect of high-frequency modes, such as the Lamb waves resonance or the effect of varying the thermal forcing profile \citep[e.g.][]{Farhat_etal_2023,Laskar_etal_2023}.
However, the 24-hour equilibrium underlines the importance of studying the combined effect of gravitational and thermal tides in more detail.

We have also shown that the tidal equilibria strongly depend on the orbital eccentricity. In particular, it gives rise to an asymmetry between the sub- and super-synchronous thermal equilibrium states.
High eccentricities also enhance the effect of gravitational tides and hence increase the critical semi-major axis that destabilises the asynchronous equilibria provided by thermal tides.
On the other hand, the eccentricity increases the number of spin-orbit resonances where the rotation can be trapped, which also helps develop more temperate climates \citep[e.g.][]{Dobrovolskis_2007, Valente_Correia_2022}.

Another striking result is that stable nonzero obliquities are a possible outcome of the tidal equilibria.
High obliquities promote a more balanced distribution of the stellar flux on the planet surface and thus extend the size of the habitable areas \citep[e.g.][]{Armstrong_2014, Colose_2019, Dobrovolskis_2021}.
In this paper, we did not explore the obliquity equilibria in great detail because other effects that we did not include in our model, such as core-mantle friction \citep[e.g.][]{Correia_etal_2003, Correia_2006, Cunha_etal_2015} and planetary perturbations \citep[e.g.][]{Laskar_Robutel_1993, Correia_Laskar_2001, Correia_Laskar_2003I, Saillenfest_etal_2019, Su_Lai_2022b} can also play an important role in the obliquity evolution.
Therefore, these additional effects should be taken into account in future studies on this topic.

The different tidal equilibria for the spin mainly depend on the orbital eccentricity, the rheology of the mantle, and the composition of the atmosphere.
The eccentricity is a parameter that can already be accessed for planets detected with the radial-velocity method \citep[e.g.][]{Sagear_Ballard_2023} and with TTVs \citep[e.g.][]{Agol_etal_2021}.
For the rheology, we expect that Solar System data will provide more constraints on the deformation and dissipation of rocky planets \citep[e.g.][]{Efroimsky_Lainey_2007, Farhat_etal_2022}.
We also expect that space telescopes such as JWST and ARIEL \citep{Edwards_Tinetti_2022} will be able to provide more insight on the composition of the atmospheres.
Inversely, if we are able to measure the spin of a given planet, we can put constraints on the unknown parameters, as we did for Venus (Fig.~\ref{Fig:3.1}).
Although not possible at present, a new generation of instruments, such as the ELT-METIS integral field spectrograph \citep{Quanz_etal_2015, Brandl_etal_2021}, will combine high-contrast imaging with high-resolution spectroscopy and allow for the determination of the spin state of these planets.

%Although we observe that the semi-major axis and eccentricity play an important role in the asynchronous equilibria, the radius of the planet, and subsequently the mass, also has an impact. Fig \ref{Fig:4.4}, shows the threshold of synchronicity and asynchronicity for an Earth-twin planet with two different atmospheres, however, if we increase the mass of the planet the critical semi-major axis is higher. Indeed, when the mass increases the radius also increases (assuming an Earth density). Because the amplitude of the gravitational tides, $\At$ is proportional to $\Rp^6$, higher than the amplitude of the thermal atmospheric tides, $\At_a \propto R^3$, it is stronger, pushing away the critical semi-major axis. So an additional study could be made in the future to understand how much the mass of the planet can influence the emergence of asynchronous states.

\begin{acknowledgements}
We are grateful to the referee Emeline Bolmont for helpful comments.
This work was supported by COMPETE 2020 and by FCT - Funda\c{c}\~ao para a Ci\^encia e a Tecnologia, I.P., Portugal, through the projects
SFRH/BD/137958/2018,
GRAVITY (PTDC/FIS-AST/7002/2020),
PHOBOS (POCI-01-0145-FEDER-029932), 
ENGAGE SKA (POCI-01-0145-FEDER-022217), and
%Enabling Green E-science for the SKA Research Infrastructure (ENGAGE SKA), 
CFisUC (UIDB/04564/2020 and UIDP/04564/2020, with DOI identifiers 10.54499/UIDB/04564/2020 and 10.54499/UIDP/04564/2020, respectively).
It was also supported by the French Agence Nationale de la Recherche (AstroMeso ANR-19-CE31-0002-01) and by the European Research Council (ERC) under the European Union’s Horizon 2020 research and innovation program (Advanced Grant AstroGeo-885250). 
We acknowledge the Laboratory for Advanced Computing at University of Coimbra (\href{https://www.uc.pt/lca}{https://www.uc.pt/lca}) for providing the resources to perform the numerical simulations.
\end{acknowledgements}

\bibliographystyle{aa}
\bibliography{references}

\end{document}